\documentclass[published]{JHEP3}

\JHEP{ }

\usepackage{cite}
\usepackage{epsf}
\usepackage{latexsym}
\usepackage{epsfig,multicol}

\def\preal{{\rm Re\,}}
\def\pim{{\rm Im\,}}

\def\yzero{\smash{\hbox{$y\kern-4pt\raise1pt\hbox{${}^\circ$}$}}}
\def\p{\partial}
\def\a{\alpha}
\def\b{\beta}
\def\g{\gamma}
\def\d{\delta}
\def\beq{\begin{equation}}
\def\eeq{\end{equation}}
\def\beqa{\begin{eqnarray}}
\def\eeqa{\end{eqnarray}}
\def\Om{\Omega}
\def\om{\omega}
\def\th{\theta}
\def\vt{\vartheta}
\def\vphi{\varphi}
\def\-{\hphantom{-}}
\def\ov{\overline}
\def\s2{\frac{1}{\sqrt2}}

\def\oh{\frac{1}{2}}
\def\beq{\begin{equation}}
\def\eeq{\end{equation}}
\def\beqa{\begin{eqnarray}}
\def\eeqa{\end{eqnarray}}

\def\Tr{{\rm Tr \,}}

\def\IF{\relax{\rm I\kern-.18em F}}
\def\II{\relax{\rm I\kern-.18em I}}
\def\IP{\relax{\rm I\kern-.18em P}}
\def\IC{\relax\hbox{\kern.25em$\inbar\kern-.3em{\rm C}$}}
\def\IR{\relax{\rm I\kern-.18em R}}

\def\cn{{\cal N}}
\def\cam{{\cal M}}

\def\Dsl{\,\raise.15ex\hbox{/}\mkern-13.5mu D} 
\def\IZ{Z\kern-.4em  Z}

\def\inte{{\bf Z}}
\def\cpx{{\bf C}}
\def\real{{\bf R}}
\def\OR{\Omega {\cal R}}
\def\R{{\cal R}}

\def\ce{{\cal E}}
\def\cl{{\cal L}}
\def\lam{\lambda}

\def\D{\Delta}

\def\raw{\rightarrow}

\def\lraw{\leftrightarrow}
\def\eps{\epsilon}
\def\k{\kappa}
\def\sig{\sigma}
\def\Sig{\Sigma}

\def\ti{\times}

\def\cw{{\cal W}}

\newbox\pippobox

\title{Yukawa couplings in intersecting D-brane models}

\author{D.~Cremades, L.~E.~Ib\'a\~nez and  F.~Marchesano \\
 	Departamento de F\'{\i}sica Te\'orica C-XI
	and Instituto de F\'{\i}sica Te\'orica  C-XVI,\\
	Universidad Aut\'onoma de Madrid,
	Cantoblanco, 28049 Madrid, Spain.
}
\received{\today}               
\revised{}
\accepted{}               

\preprint{\hepth{0302105}}
\preprint{FTUAM-03/2 IFT-UAM/CSIC-03-06}

\abstract{We compute the Yukawa couplings among chiral
fields in toroidal Type II compactifications with
wrapping D6-branes intersecting at angles. 
Those models can yield realistic standard model
spectrum living at the intersections. The Yukawa couplings
depend both on the K\"ahler and open string moduli
but not on the complex structure. They  
arise from worldsheet instanton corrections and
are found to be given by products of complex Jacobi theta 
functions with characteristics.  
The Yukawa couplings for a  particular 
intersecting brane configuration yielding the 
chiral spectrum of the MSSM are computed
as an example. We also show how our methods can be 
extended to compute Yukawa couplings on certain classes 
of elliptically fibered  CY manifolds which are mirror to
complex cones over del Pezzo surfaces. We find that 
the Yukawa couplings in intersecting D6-brane 
models have a mathematical interpretation in the
context of homological mirror symmetry. 
In particular, the computation of such Yukawa couplings 
is related to the construction of Fukaya's category 
in a generic symplectic manifold. }

\keywords{ D-branes, Yukawa couplings, Mirror symmetry,  String Phenomenology}

\begin{document}

\section{Introduction}

Since the middle eighties  there has been a lot of 
work and effort devoted to relate string theory to the observed  
world. In particular,  superstring phenomenology aims at
 obtaining  the  observed  low
energy physics as an effective theory of an 
string-based model. There is, however, still a big gap between theory and 
experiment. One of the latest proposals regarding a construction of 
realistic superstring vacua is based on the so-called Intersecting Brane 
World scenario. This scenario naturally involves the brane-world idea, 
where gauge interactions are confined in some lower dimensional 
submanifold ({\it brane}) of a larger manifold ({\it bulk}) 
where gravitational interactions do also propagate. In addition, it 
incorporates a simple mechanism to obtain one of the most important 
properties of the Standard Model (SM) of particle physics, which is 
chirality. Indeed, two intersecting D-branes will yield a massless 
chiral fermion localized at their intersection \cite{bdl}. The presence 
of these two appealing ingredients in the stringent context of string 
theory makes this proposal rather promising from the phenomenological 
point of view.

Indeed, in the last two years, the intersecting D-brane approach has 
been particularly successful  in the building-up of 
semi-realistic string theory models
\cite{bgkl,afiru,afiru2,bkl,imr,bklo,csu,bailin,hon,cim1,cim2,koko,
cimD5,CYberlin,ellis,local,bgo,cps}. 
Most models are toroidal or orbifold (orientifold) compactifications
of Type II string theory with  Dp-branes wrapping intersecting cycles
on the compact space. At the different brane intersections
live chiral fields to be identified with SM fermions.
There is a natural origin for the replication of quark-lepton generations
since the Dp-branes wrapping a compact space typically 
intersect a multiple number of times.

Several phenomenologically 
interesting results  have been obtained, such as
the construction of specific models \cite{imr} yielding just the 
chiral spectrum of the Standard Model. 
This class of constructions present and interesting   
structure of global $U(1)$ symmetries that arise from the underlying
 D-brane configuration. In these theories baryon number
is gauged, insuring proton stability to all orders in
perturbation theory. 
Another important issue has been the achievement of chiral
supersymmetric vacua by means of intersecting D-branes \cite{csu,bgo},
 which moreover allow to construct configurations with three families
of quark and leptons.
The stability of such theories and their potential lift to M-theory
makes them of central interest also from the theoretical viewpoint.
Intersecting branes have also inspired the more exotic constructions
named q-SUSY theories, which basically are
non-supersymmetric theories where quadratic divergences appear
only at two loops \cite{cim1,cim2}.

One of the most attractive features of the brane-world scenario is the 
possibility of weakening gravitational interactions by considering they 
propagate in large extra dimensions where SM interactions do not 
\cite{add}. Realistic scenarios where such mechanism could work were 
constructed in \cite{cimD5}, involving intersecting D-branes at orbifold 
singularities. The same mechanism, but now on the broader 
context of Calabi-Yau geometry has been developed in \cite{local}.
In fact, almost all the above constructions, and in particular 
the compactifications 
yielding a realistic spectrum, have been performed either in toroidal 
geometries or in orbifold/orientifold quotients of these. The 
generalization of realistic constructions to more complicated geometries 
as, e.g., general Calabi-Yau manifolds, has been adressed in 
\cite{CYberlin,local}.

The application of intersecting branes to string phenomenology is not,
though, restricted to SM physics. In \cite{raul} it was proposed a
cosmological scenario where inflation arises from intersecting brane 
dynamics. This idea has been pursued in \cite{berlinInf,tye,marta},
yielding new interesting scenarios.

We thus see that  intersecting brane
 configurations  provide promising setups where to accommodate semi-realistic 
low-energy physics. Considering this, one may wonder how close can we 
get to, say, a string compactification providing the SM as a low-energy 
effective theory. As we know, the Standard Model is not a bunch of 
chiral fermions with appropriate quantum numbers, but an intricate 
theory with lots of well-measured parameters. The next step in this 
quest, then, might be checking whether we can reproduce some of these 
finer data defining the SM.

The main purpose of this paper is to address the computation of 
Yukawa couplings in the context of intersecting brane worlds. As  
advanced in \cite{afiru2}, those arise from open string worldsheet 
instantons that connect three D-brane intersections, in such a way that 
the open string states located there have suitable Lorentz and gauge 
quantum numbers to build up an invariant.
%
\EPSFIGURE{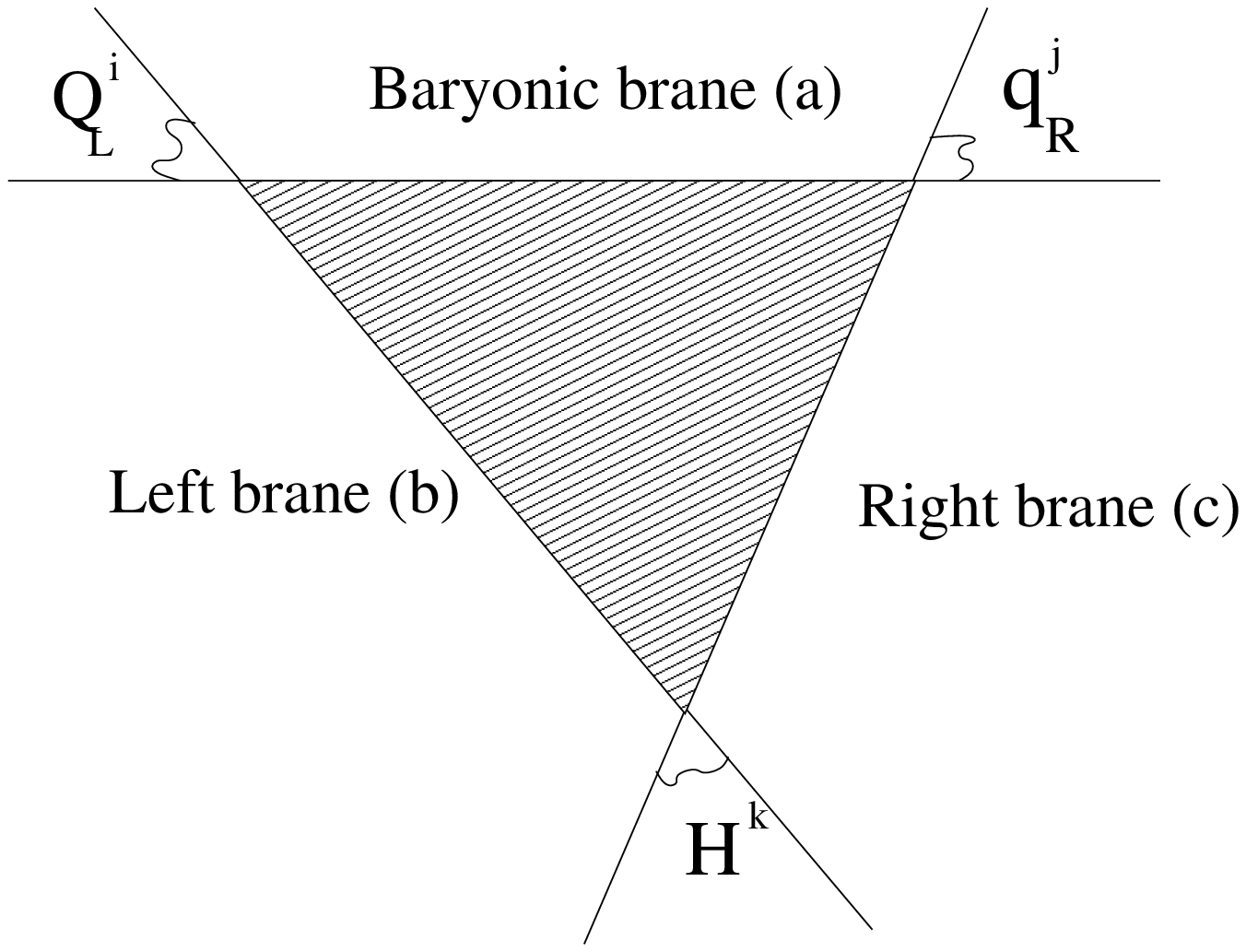, width=3.2in}
{\label{yukis2}
Yukawa coupling between two quarks of opposite chirality and a Higgs boson.}
%
A paradigmatic example is presented in figure \ref{yukis2}, where two 
quarks of opposite chirality couple to a Higgs boson. The
worldsheet connecting the corresponding intersection has the topology of 
a disc, and involves three different boundary states. 
These are, in a target space perspective, branes $a$, $b$ 
and $c$, whose intersections localize the matter particles. Given such 
boundary conditions, there will exist an infinity of worldsheets satisfying 
them. In order to compute the instanton correction to our effective theory,
 however, we must concentrate on minimal action worldsheets
that satisfy the classical equations of motion, and sum over topologically 
different sectors (see below). Each of these will contribute to the Yukawa 
coupling as something proportional to $exp(-S_{cl})$ so, na\"{\i}vely, we 
would expect it to be weighted by exp($-A_{abc}$), where $A_{abc}$ is the 
worldsheet area.

As a result, Yukawa couplings will depend both on the D-brane positions 
and on the geometry of the underlying compact space. In terms of low energy 
quantities, these are characterized by open and closed string moduli 
v.e.v.'s. We have computed such dependence explicitly in the simple case 
of D-branes wrapping factorizable cycles on flat tori of arbitrary dimension. 
We have considered, as well, how these Yukawa are affected when more general 
configurations including non-vanishing B-field and Wilson lines are 
included. We find that Yukawa couplings have
simple  expressions in terms of complex theta functions 
with characteristics or, ultimately, in terms of multi-theta functions.
This is somewhat analogous to the structure found for
the Yukawa couplings of toroidal heterotic orbifold models
as computed in \cite{orbif}. 
No complex structure dependence appears in our Yukawa  
couplings, and the wrapping numbers of the D-branes configuration appear 
only through their intersection numbers on each subtorus.
Thus given those intersection numbers one can immediately write
down the Yukawa couplings for any toroidal intersecting 
brane model, there is no explicit dependence on the particular
wrapping numbers of the given brane configuration.
These general facts are illustrated by an explicit example
presented below, based on D6-branes wrapping cycles on (an orientifold of)
$T^2\times T^2\times T^2$ and with the chiral spectrum
of the minimal supersymmetric standard model (MSSM).
Explicit results for the Yukawa couplings are presented in
this example, which reproduce the leading effect of
one generation of quarks and leptons being much heavier  
than the first two.

The formulation we present can also be applied to 
some non-toroidal CY compactifications, 
i.e., certain elliptically fibered manifolds.
 In particular, the general class of elliptically fibered non-compact
CY manifolds considered in ref.\cite{local}  which are 
mirror to complex cones over del Pezzo surfaces.
Configurations of D6-branes wrapping cycles on the
original CY have mirror configurations corresponding to
D3-branes located at those singularities. 
A particularly simple case is that of D3-branes sitting at a
$\inte_3$ singularity which is mirror to D6-branes wrapping
cycles on an elliptically fibered non-compact CY manifold.
Using our methods we compute the general form of Yukawa couplings 
in the wrapping D6-brane configuration and check that they match
the Yukawa couplings known for D3-branes at a $\inte_3$ 
singularity. 

The structure of this paper is as follows.
In the next section we discuss the r\^ole of world-sheet instantons 
in the computation of Yukawa couplings in general intersecting
brane configurations. In section 3 we explicitly compute the
Yukawa couplings for intersecting configurations of
D-branes wrapping cycles in a toroidal (orientifold) 
compactification. We also analyze the more general case in
which a B-field and Wilson line backgrounds are added.
Explicit expressions in terms of products of
Jacobi complex theta functions with characteristics 
are given. In section 4 we discuss an explicit example
yielding the chiral spectrum of the MSSM and 
provide the corresponding Yukawa couplings.
A CY example (mirror to the case of D3-branes sitting
on a $\cpx^3/\inte_3$ singularity) is briefly studied in section 5.
We check there that the Yukawa couplings obtained 
from our method match the known results of the mirror.
After performing our computations of section 3, we realized
that they were intimately related to some previous work
in the mathematical literature, in the very different 
context of homological mirror symmetry.
Section 6 is devoted to briefly discuss such connection and,
in particular, to show how computation of Yukawa couplings 
in intersecting D-brane models can be translated to
the computation of Fukaya category in a generic symplectic 
manifold.
Final comments and conclusions are left for section 7.

\section{Intersecting brane models and Yukawa couplings}

In this section we study intersecting D-brane models from a general 
viewpoint\footnote{For a nice recent review on these topics see 
\cite{angel03}.}, reviewing some previous work on the field and collecting 
the necessary information for addressing the problem of Yukawa couplings. 
Most part of the effort on constructing phenomenologically appealing 
intersecting brane configurations has centered on simple toroidal and on 
orbifold/orientifold compactifications. However, main issues as, e.g., 
massless chiral spectrum and tadpole cancellation conditions, are of 
topological nature and thus easily tractable in more general 
compactifications where the metric may not be known explicitly. Following 
this general philosophy, we will introduce Yukawa couplings as arising from
 worldsheet instantons in a generic compactification. Although the specific
 computation of these worldsheet instantons needs the knowledge of the target
 space metric, many important features can be discussed at this more general 
level. In the next section we will perform such explicit computation in the 
simple case of toroidal compactifications, giving a hint of how these 
quantities may behave in a more general setup.

\subsection{D-branes wrapping intersecting cycles}

Consider type IIA string theory compactified on a six dimensional manifold 
$\cam$. \footnote{Thorough this section we will we working in the large 
volume limit of compactification.}
The building blocks of an intersecting brane configuration will be 
given by D6-branes filling four-dimensional Minkowski space-time and wrapping 
internal homology 3-cycles of $\cam$. 
\footnote{In general, we could conceive
 constructing chiral four-dimensional models from type IIA or type IIB  
intersecting branes, other than D6-branes, that sit on singular orbifold 
fixed points. Indeed, some of these models have been constructed in 
\cite{afiru,afiru2,bailin,hon,cimD5}. However, as emphasized  in 
\cite{local}, these configurations can be related to intersecting
 D6-branes either by blowing up orbifold fixed points or by means of 
mirror symmetry.} A specific configuration will thus consist of $K$ stacks of 
D6-branes, each stack $\a$ containing $N_\a$ coincident D6-branes whose
 worldvolume is given by ${\bf M_4} \ti \Pi_\a \subset {\bf M_4} \ti \cam$,
 where $[\Pi_\a] \in H_3(\cam,\inte)$ is the corresponding homology class 
of such 3-cycle. The gauge theory arising from open string degrees of freedom 
will be localized on such D6-brane worldvolumes, giving rise to a total gauge
 group $\prod_\a U(N_\a)$. 

In addition, there will be some open strings modes arising from strings 
stretched between, say, stacks $\a$ and $\b$. In case the corresponding 
3-cycles $\Pi_\a$ and $\Pi_\b$ intersect at a single point in the compact
 space $\cam$, the lowest open string mode in the R sector will 
correspond to a chiral fermion localized at the four-dimensional 
intersection of $\a$ and $\b$,
 transforming in the bifundamental representation of $U(N_\a) \ti U(N_\b)$ 
\cite{bdl}. Notice that, since $\cam$ is a compact manifold, $\Pi_\a$ and 
$\Pi_\b$ will generically intersect several times. The number of such 
localized chiral fermions is given by the number of intersections 
$\# (\Pi_\a \cap \Pi_\b)$. This number is not, however, a topologically 
invariant quantity. Such invariant is constructed from subtracting the 
number of right-handed chiral fermions to left-handed ones, after what we 
obtain the intersection number $I_{\a\b} \equiv [\Pi_\a] \cdot [\Pi_\b]$, 
which give us the {\em net} number of chiral fermions in the $\a\b$ sector.

As we are considering $\cam$ to be compact, any configuration should satisfy 
some consistency conditions related to the propagation of Ramond-Ramond 
massless closed string fields on $\cam$. These are the RR tadpole cancellation
conditions which  
 require the total RR charge of the configuration
 to vanish. In our case, the charge of a D6-brane $\a$ under the RR 7-form 
$C_7$ is classified by its associated 3-cycle homology cycle $[\Pi_\a]$. 
Hence, RR tadpoles amount to imposing that the sum of homology classes add 
up to zero \cite{angel00}
\beq
\sum_\a N_\a [\Pi_\a] = 0.
\label{tadpoles}
\eeq
Additional RR sources may appear in general, such as O6-planes arising in 
orientifold compactifications or NS-NS background fluxes. Each of theses 
objects will have an associated 3-cycle homology class, so that RR conditions 
will be finally expressed again as the vanishing of the total homology class. 
It can be easily seen that RR tadpole conditions directly imply the 
cancellation of non-abelian $SU(N_\a)^3$ anomalies. They also imply, by the 
mediation of a generalized Green-Schwarz mechanism \cite{sagno1,iru}, 
the cancellation of mixed non-abelian 
and gravitational anomalies \cite{afiru,imr,csu,angel02}.

So far, we have not imposed any particular constraint on our manifold $\cam$,
 except that it must be compact so that we recover four-dimensional gravity 
at low energies. We may now require the closed string sector to be 
supersymmetric. This amounts to impose that $\cam$, seen as a Riemannian 
manifold with metric $g$, has a holonomy group contained in $SU(3)$. Now, 
such a manifold can be equipped with a complex structure $J$ and a 
holomorphic volume 3-form $\Om_c$ which are invariant under the holonomy 
group, i.e., covariantly constant. This promotes $\cam$ to a Calabi-Yau 
three-fold, or ${\bf CY_3}$. \footnote{For reviews on Calabi-Yau geometry 
see, e.g., \cite{tristan,joyce}.} Moreover, $g$ and $J$ define a K\"ahler 
2-form $\om$ which satisfies the following relation with the volume form
\beq
\frac{\om^3}{3!} = \left( \frac i2\right)^3 \Om_c \wedge \ov \Om_c.
\label{omegas}
\eeq
Given a real 3-form $\Om$ normalized as this, we can always take 
$\Om_c = e^{i\th} \Om$ for any phase $\th$ as a solution of (\ref{omegas}). 
The K\"ahler form $\om$ can also be complexified to $\om_c$, by addition of
 a non-vanishing $B$-field. Both $\Om$ and $\om$ will play a central r\^ole
 when considering the open string sector.

Notice, however, that we have not imposed Hol$(\cam)$ to be {\em exactly} 
$SU(3)$.
\footnote{This is an important phenomenological restriction when 
considering, e.g., perturbative heterotic compactifications. 
This is not longer the case 
on Type II or Type I theories, where bulk supersymmetry can be further 
broken by the presence of D-branes.} 
We may consider, for instance, 
$\cam = T^2 \ti {\bf K3}$, whose holonomy group is contained in $SU(2)$. 
In this case, there is not a unique invariant 3-form but two linearly 
independent ones, both satisfying (\ref{omegas}). In general, the number 
of (real) covariantly constant 3-forms of $\cam$ satisfying (\ref{omegas}) 
indicates the amount of $D = 4$ supersymmetries preserved under 
compactification. In a ${\bf CY_3}$ in the strict sense, with 
Hol$(\cam) = SU(3)$, the gravity sector yields $D = 4$ $\cn = 2$ under 
compactification, and this fact is represented by the 
existence of a unique complex volume 
form $e^{i\th} \Om$. Indeed, $\th$ parametrizes the $U(1)$ of $\cn = 1$ 
superalgebras inside $\cn = 2$ \cite{douglas}. Correspondingly, 
compactification on $T^2 \ti {\bf K3}$ yields a $D = 4$ $\cn = 4$ gravity 
sector. There are some other consequences when considering manifolds of 
lower holonomy. For instance, Hol$(\cam) = SU(3)$ implies that 
$b_1(\cam) = 0$ \cite{tristan}, while this might not be the case for 
lower holonomy, as the example $T^2 \ti {\bf K3}$ shows.

Let us now turn to the open string sector, represented by type IIA D6-branes.
 Intuitively, a dynamical object as a D-brane will tend to minimize its 
tension while conserving its RR charges. In our geometrical setup, this 
translates into the minimization of Vol($\Pi_\a$) inside the homology class 
$[\Pi_\a]$. A particular class of volume-minimizing objects are calibrated 
submanifolds, first introduced in \cite{hl}. The area or volume of such 
submanifolds can be computed by integrating on its $p$-volume a (real) 
closed $p$-form, named calibration, defined on the ambient space $\cam$. 
Both $\Om$ and $\om$ are calibrations in a ${\bf CY_3}$. Submanifolds 
calibrated by $\Om$ are named special Lagrangian \cite{joyce} while those 
calibrated by $\om$ are holomorphic curves. Being a 3-form, $\Om$ will 
calibrate 3-cycles where D6-branes may wrap. A 3-cycle $\Pi_\a$ calibrated 
by Re($e^{i\th}\Om$) will have a minimal volume on $[\Pi_\a]$, given by 
Vol($\Pi_\a$) $ = \int_{\Pi_\a}$ Re($e^{i\th}\Om$), and will be said to
 have phase $\th$. We can also characterize such calibration condition by
\beq
\om|_{\Pi_\a} \equiv 0 \quad {\rm and} \quad 
{\rm Im} (e^{i\th}\Om)|_{\Pi_\a} \equiv 0.
\label{cali2}
\eeq
The middle-homology objects that satisfy the first condition in 
(\ref{cali2}) are named Lagrangian submanifolds and, although they 
are not volume minimizing, play a central r\^ole in symplectic geometry. 

A D6-brane whose 3-cycle $\Pi_\a$ wraps a special Lagrangian (sL) submanifold
 does not only minimize its volume but, as shown in \cite{bbs}, also preserves
 some amount of supersymmetry. Being BPS stable objects of type IIA theory, 
it seems natural to consider D6-branes wrapping sL's as building blocks of 
our intersecting brane configurations. 

\subsection{The r\^ole of worldsheet instantons}

Being a BPS soliton of type IIA theory, a stack of $N_\a$ D6-brane wrapping 
a special Lagrangian submanifold $\Pi_\a$ will yield a $U(N_\a)$ 
Supersymmetric Yang-Mills theory on its worldvolume. By simple dimensional 
reduction, the (inverse) 
gauge coupling constant can be seen to be proportional to 
Vol$(\Pi_\a)$, which can be computed by integrating Re($e^{i\th}\Om$) on 
the corresponding homology cycle. A natural question is which amount of 
SUSY the D6-brane effective field theory will have by dimensional reduction 
down to $D =  4$. The precise amount is again given by the number of 
independent real volume forms Re$(e^{i\th}\Om)$ that calibrate the 3-cycle.
We may, however, seek for a more topological alternative method.

McLean's theorem \cite{McLean} states that the moduli space of deformations
 of a sL $\Pi_\a$ is a smooth manifold of (unobstructed) real dimension 
$b_1(\Pi_\a)$. String theory complexifies this space, by adding the 
$b_1(\Pi_\a)$ Wilson lines obtained from the gauge field 
$U(N_\a)$ living on the worldvolume of the stack $\a$ 
In the low energy $D=4$ theory, this will translate into $b_1(\Pi_\a)$ 
massless complex scalar fields in the adjoint of $U(N_\a)$. 
Being in a supersymmetric theory, these fields will yield the scalar 
components of $D = 4$ supermultiplets. 

Let us consider the case $\cn = 1$ (for a clear discussion on this see, e.g., 
\cite{kachru}). Here we find $b_1(\Pi_\a)$ chiral multiplets $\phi_j$ in the 
adjoint of $U(N_\a)$. Now, we may also seek to compute the superpotential 
$W$ of such $\cn = 1$ theory, which is a function on these chiral fields. 
By standard $\cn = 1$ considerations, this superpotential cannot be 
generated at any order in $\a'$ perturbation theory, in accordance with the 
geometrical result of \cite{McLean}. Indeed, such 
superpotential will be generated non-perturbatively by worldsheet instantons 
with their boundary in $\Pi_\a$.

Superpotentials generated non-perturbatively by worldsheet instantons were 
first considered in closed string theory \cite{dsww}, while the analogous 
problem in type IIA open string has been recently studied in \cite{vafa,kklm},
 in the context of open string mirror symmetry. The basic setup considered is 
a {\it single} D6-brane  wrapping a sL $\Sig $  in a ${\bf CY_3}$, with 
$b_1(\Sig) > 0$. Worldsheet instantons are constructed by considering all 
the possible embeddings of a Riemann surface ${\cal S}_g$ with arbitrary 
genus $g$ on the target space $\cam$ and with boundary on $\Sig$. In order 
to be topologically non-trivial, this boundary must be wrapped on the 
1-cycles $\g_j$ that generate $b_1(\Sig)$, thus coupling naturally to the 
corresponding chiral multiplets. Moreover, in order to deserve the name 
instanton, this euclidean worldsheet embedding must satisfy the classical 
equations of motion. This is guaranteed by considering embeddings which are 
holomorphic (or antiholomorphic) with respect to the target space complex 
structure, plus some extra constraints on the boundary (Dirichlet conditions).
 In geometrical terms, this means that {\em worldsheet instantons 
must be surfaces calibrated by the K\"ahler form $\om$}. Calibration theory 
then assures the area minimality given such boundary conditions, which is 
what we would expect from na\"{\i}ve Nambu-Goto considerations. As a general 
result, it is found that the superpotential of D6-brane theories is entirely 
generated by instantons with the topology of a disc, while higher-genus 
instantons correspond to open string analogues of Gromov-Witten invariants.

So we find that, in case of $\cn = 1$ D6-branes on a ${\bf CY_3}$, great deal 
can be extracted from calibrated geometry of the target space $\cam$. Whereas
 the gauge kinetic function $f_{ab}$ can be computed by evaluating the volume
 form $\Om$ on the worldvolume $\Sig$ of the brane, the superpotential can be
 computed by integrating the K\"ahler form $\om$ on the holomorphic discs 
with boundary on $\Sig$. The former only depends on the homology class 
$[\Sig] \in H_3(\cam, \inte)$, and in the case of toroidal compactifications 
they have been explicitly computed in \cite{cim1}. The latter, on the 
contrary, is given by a sum over the relative homology class 
$H_2^D(\cam,\Sig)$, that is, the classes of 2-cycles on $\cam$ with boundary 
on $\Sig$ (the superscript $D$ means that we only consider those 2-cycles with
 the topology of a disc). Notice that, $\cam$ being compact, the disc 
instantons may wrap multiple times. Although in principle one may need the 
knowledge of the metric on $\cam$ in order to compute both, much can be 
known about the form of the superpotential by considerations on Topological 
String Theory.
For our purposes, we will contempt to stress two salient features:
\begin{itemize}

\item The superpotential depends on the target space metric only by means 
of K\"ahler moduli, and is independent of the complex structure \cite{quintic}.

\item If we see those K\"ahler moduli as closed string parameters, the 
dependence of the superpotential is roughly of the form
\beq
W = \pm \sum_{n=1}^\infty {e^{-n\Phi} \over n^2},
\label{super}
\eeq
 where $n$ indexes the multiple covers of a disc with same boundary 
conditions, and $\Phi$ stands for the open string chiral superfield 
\cite{ov}.
 The $\pm$ sign corresponds to holomorphic and antiholomorphic maps, 
respectively.

\end{itemize}

Given these considerations, is easy to see that no superpotential will be 
generated for a chiral superfield associated to a 1-cycle $\g$ of $\Sig$ 
which is also non-contractible in the ambient space $\cam$, since no disc 
instanton exist that couples to such field. Notice that this could never 
happen in a ${\bf CY_3}$ on the strict sense, since in this case 
$b_1(\cam) = 0$. In manifolds with lower holonomy, however, it may 
well be the case that $b_1(\cam) > 0$, and so a D6-brane could have in its 
worldvolume a complex scalar not involved in the superpotential (\ref{super}). 
We expect such scalars to give us the scalar content of the vector 
supermultiplet, 
thus indicating the degree of supersymmetry on the worldvolume effective 
theory. This seems an alternative method for computing the amount of 
supersymmetry that such a D6-brane preserves. A clear example of the above 
argument is constituted by $\cam = T^2 \ti T^2 \ti T^2$ and $\Sig$ a 
Lagrangian $T^3$ (the so called factorizable branes considered in the next 
section). Here, each of the three independent 1-cycles on $\Sig$ is 
non-contractible in $\cam$, so our SYM theory will yield three complex 
scalars not involved in the superpotential. But these scalar fields fill in 
the precise content of a $D = 4$ $\cn = 4$ vector multiplet, which is the 
amount of SUSY those branes preserve.

\subsection{Yukawa couplings in intersecting D-brane models}

\begin{figure}[ht]
\begin{center}
\begin{tabular}{ll}
\\
\hskip -0.5cm
\epsfig{file=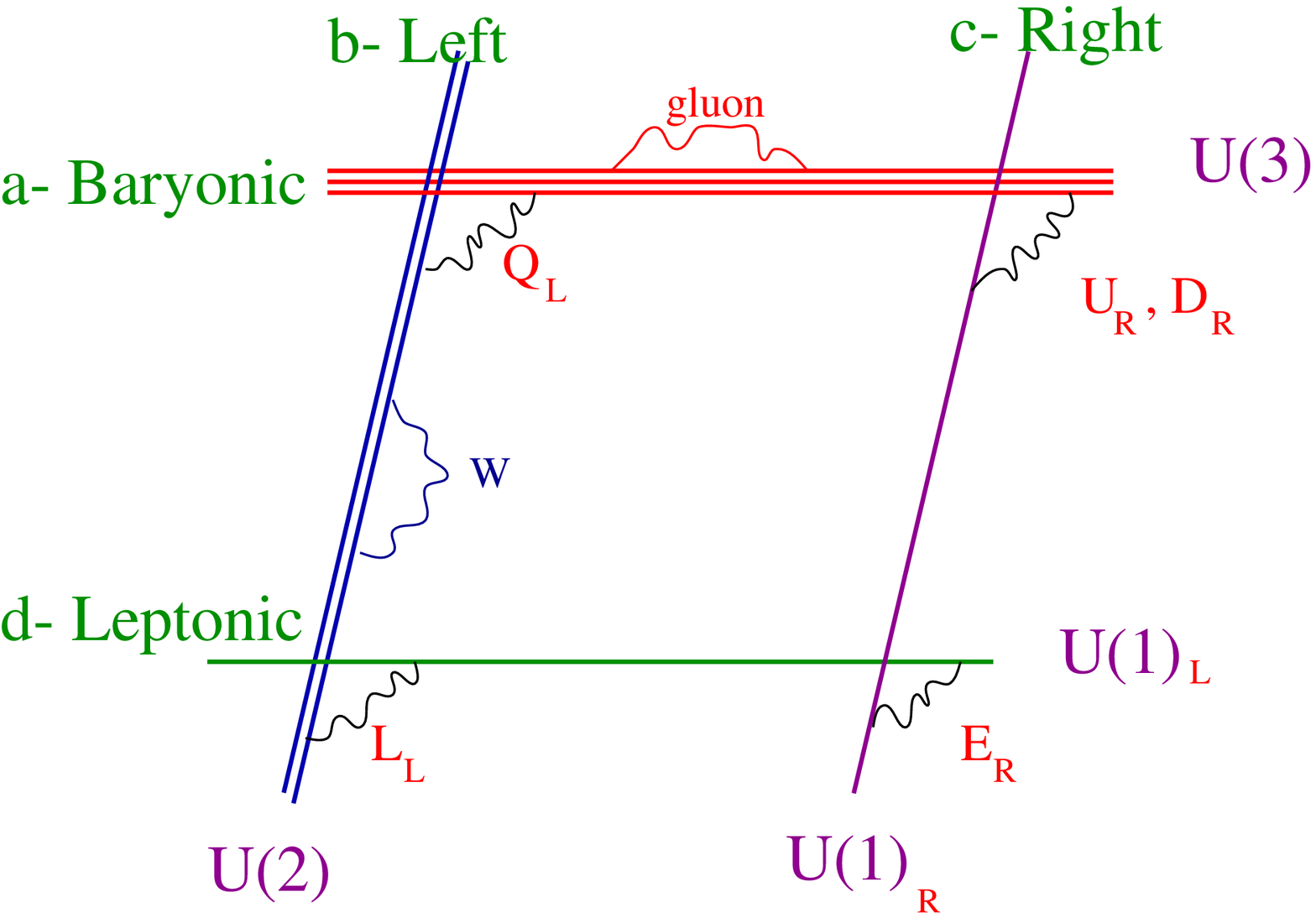, height=5.5cm} &
\epsfig{file=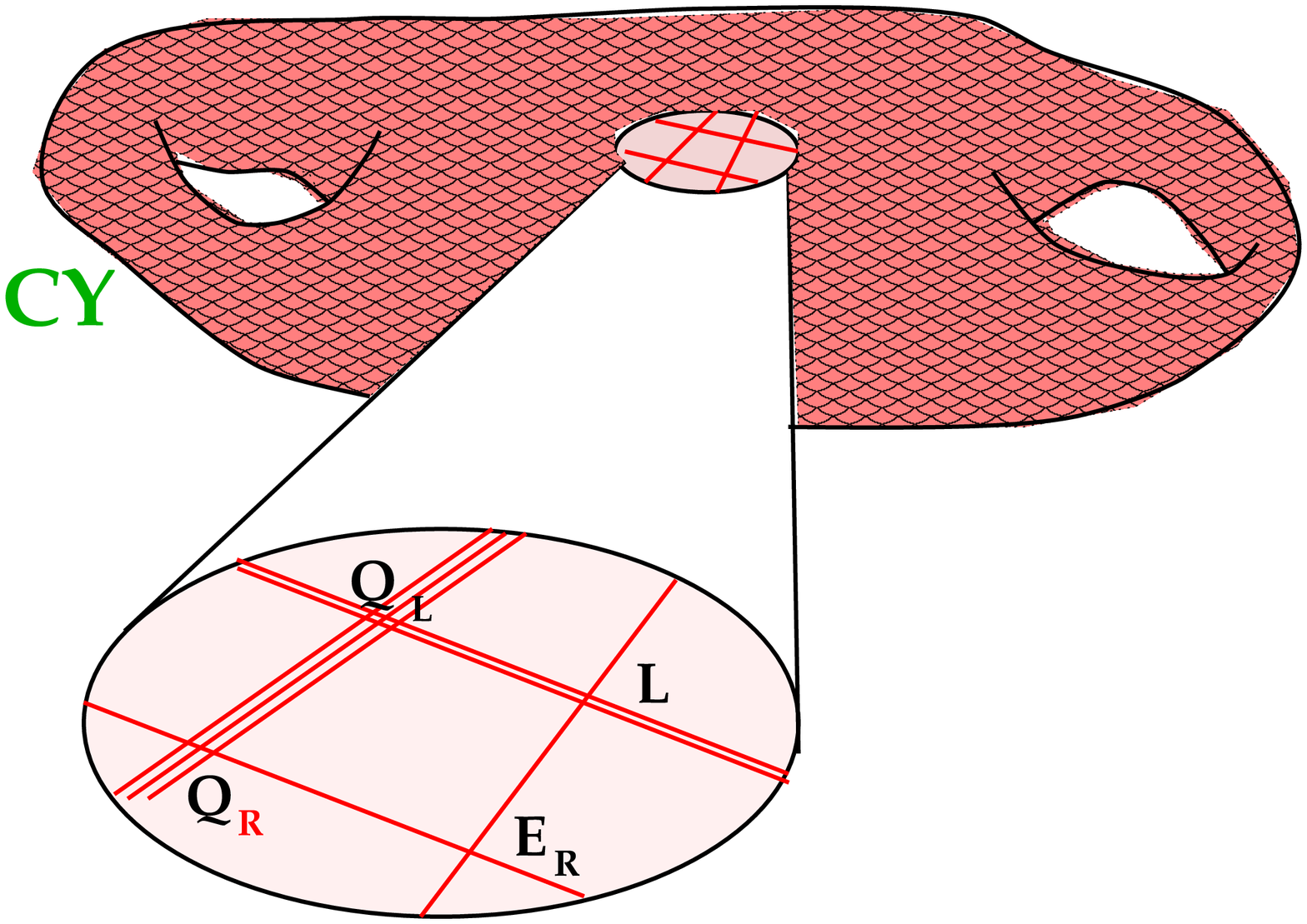, height=5.5cm}\\
\hskip 3.1truecm
{\small (a)}            &
\hskip 3.65truecm
{\small (b)}\\
\end{tabular}
\end{center}
\caption{\small{The Standard Model at intersecting D-branes.
a) Four stacks of branes, {\it baryonic, left, right } and 
{\it leptonic}
are needed to get all quark and leptons at the intersections.
b) The SM branes may be wrapping cycles on a, e.g., CY manifold,
with appropriate intersections so as to yield the SM chiral spectrum.}}
\label{sm}
\end{figure}

Up to now, we have only considered superpotentials arising from one single 
stack of D6-branes. In the intersecting brane world picture we have given 
above, however, chiral matter in the bifundamental arises from the 
intersection of two stacks of branes, each with a different gauge group. It 
thus seems that, in order to furnish a realistic scenario, several stacks of 
branes are needed. In fact, given the semi-realistic model-building 
considered up to now, it seems that a minimal number of four stacks of 
branes are necessary in order to accommodate the chiral content of the 
Standard Model in bifundamentals \cite{imr}. These stacks have been named 
as {\it Baryonic (a)}, {\it Left (b)}, {\it Right (c)} and 
{\it Leptonic (d)}, in account of the global quantum numbers they carry.
 The gauge theory they initially yield is $U(3) \ti U(2) \ti U(1) \ti U(1)$,
 which arises from stack multiplicities $N_a = 3$, $N_b = 2$, $N_c = 1$ and 
$N_d = 1$. \footnote{This picture may be slightly changed in orientifold 
models, see e.g., section 4 below.} Although this yields extra abelian gauge 
factors, their gauge bosons may become massive by coupling to closed string 
RR fields, showing up in the low energy limit as global $U(1)$ symmetries 
\cite{imr}. Standard Model chiral fermions will naturally arise from pairs 
of intersecting stacks. For instance, left-handed quarks will arise from the 
intersection points of baryonic and left stacks, and so on. This scenario 
has been depicted schematically in figure \ref{sm}. For short reviews on this 
subject see \cite{reviews}.

Notice that considering a full D6-brane configuration instead of one 
single brane makes the supersymmetry discussion more involved. Although 
each of the components of the configuration (i.e., each stack of D6-branes) 
is wrapping a special Lagrangian cycle and thus yields a supersymmetric 
theory on its worldvolume, it may well happen that two cycles do not preserve 
a common supersymmetry. In a ${\bf CY_3}$ of $SU(3)$ holonomy this picture 
is conceptually quite simple. There only exist one family of real volume 
forms $\Om$ parametrized by a phase $e^{i\th}$. Two sL's $\Pi_\a$, $\Pi_\b$ 
will preserve the same supersymmetry if they are calibrated by the same real 
3-form, that is, if $\th_\a = \th_\b$ in (\ref{cali2}). In this case, a 
chiral fermion living at the intersection $\Pi_\a \cap \Pi_\b$ will be 
accompanied by a complex scalar with the same quantum numbers, filling up a 
$\cn = 1$ chiral multiplet 
\footnote{Departure from the equality of angles 
will be seen as  Fayet-Iliopoulos terms in the effective $D=4$ field theory. 
Contrary to the superpotential, these FI-terms are predicted to depend only 
on the complex structure moduli of the ${\bf CY_3}$. These aspects have been 
explored in \cite{douglas,km} in the general case, and computed from the 
field theory perspective in the toroidal case in \cite{cim1}.}. 
In manifolds 
of lower holonomy, however, there are far more possibilities, since many more
 SUSY's are involved. Consideration of such possibilities lead to the idea of
 Quasi-Supersymmetry in \cite{cim1,cim2} (see \cite{matthias} for related 
work). In order to simplify our discussion, we will suppose that all the 
branes preserve the same $\cn = 1$ superalgebra, although our results in 
the next section seem totally independent of this assumption.

It was noticed in \cite{afiru2} that, in the context of intersecting 
brane worlds, Yukawa couplings between fields living at brane intersections 
will arise from worldsheet instantons involving three different boundary 
conditions (see figure \ref{yukis3}). Let us, for instance, consider a 
triplet of D6-brane stacks and suppose them to be the Baryonic, Left and 
Right stacks, wrapping the sL's $\Pi_a$, $\Pi_b$ and $\Pi_c$ respectively. 
By computing the quantum numbers of the fields at the intersections, we 
find that fields $Q_L^i \in \Pi_a \cap \Pi_b$ can be identified with 
Left-handed quarks, $q_R^j \in \Pi_c \cap \Pi_a$ with Right-handed quarks 
and finally $H^k \in \Pi_b \cap \Pi_c$ with Higgs particles 
\footnote{Actually, in order for these fields to be properly identified 
with the SM particles is crucial that we fix both their 
{\em multiplicity} and their {\em chirality} to be the right one. 
As explained above, this imposes a topological condition on the intersection
 number, in this case $[\Pi_a]\cdot[\Pi_b] = [\Pi_c]\cdot[\Pi_a] = 3$. For
 an example on this model-building constraints see the semi-realistic model 
of section 4.}. A Yukawa coupling in SM physics will arise from a coupling 
between these three fields. In our context, such trilinear coupling will 
arise from the contribution of open worldsheet instantons with the topology 
of a disc and with three insertions on its boundary. Each of these insertions
 corresponds to an open string twisted vertex operators that changes boundary
 conditions, so that finally three different boundaries are involved in the 
amplitude. 

\EPSFIGURE{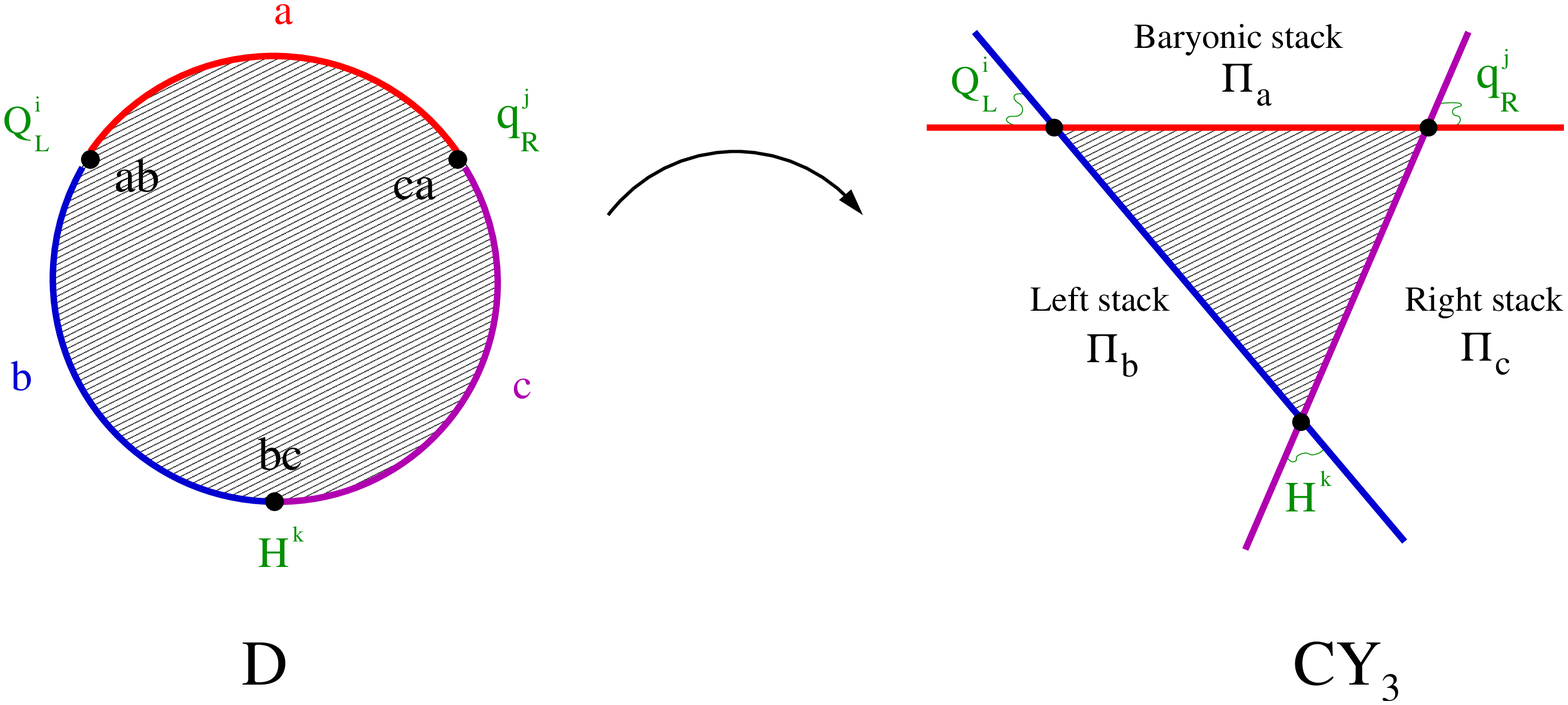, width=6in}
{\label{yukis3}
Yukawa couplings as euclidean maps from the worldsheet.}

From the target space perspective, such amplitude will arise from an 
(euclidean) embedding of the disc in the compact manifold $\cam$, with 
each vertex operator mapped to the appropriate intersection of two branes 
(generically a fixed point in $\cam$) and the disc boundary between, say, 
$ca$ and $ab$ to the worldvolume of the D6-brane stack $a$, etc. 
Such mapping $D \rightarrow \cam$ has been schematically drawn in figure 
\ref{yukis3}. 

Notice that an infinite family of such maps exist. However, only a subfamily 
satisfies the classical equations of motion, thus corresponding to true 
semiclassical instanton configurations. As expected, these correspond to 
holomorphic embeddings of $D$ on $\cam$ with the boundary conditions 
described above. Just as in the previous case of one single D6-brane, 
these instantons correspond to surfaces calibrated by the K\"ahler form 
$\om$, hence of minimal area. In the specific setup discussed in 
\cite{afiru2}, the geometry of intersecting brane worlds is reduced to 
each stack wrapping (linear) 1-cycles on a $T^2$. It is thus easy to see 
that in this case the target space worldsheet instanton has a planar 
triangular shape. This, however, will not be the general shape for a 
holomorphic curve, even in the familiar case of higher dimensional tori 
with a flat metric (see appendix A).

More concretely, we expect the Yukawa couplings between the fields  
$Q_L^i$, $q_R^j$ and $H^k$ to be roughly of the form:
\beq
Y_{ijk} = h_{qu} \sum_{\vec{n} \pm \in H_2^D(\cam,\cup_\a \Pi_\a,ijk)} 
{d_{\vec{n}} \ e^{-{A_{ijk}(\vec{n}) \over 2\pi \a^\prime}} 
e^{-2\pi i \phi_{ijk}(\vec{n})}}.
\label{yukabs}
\eeq
Here $\vec{n}$ is an element of the relative homology group 
$H_2(\cam;\cup_\a \Pi_\a, \inte)$, that is, a 2-cycle in the Calabi-Yau 
$\cam$ ending on $\cup_\a \Pi_\a = \Pi_a \cup \Pi_b \cup \Pi_c$. We 
further impose this 2-cycle to have the topology of a disc, and to connect 
the intersections $i \in \Pi_a \cap \Pi_b$, $j \in \Pi_c \cap \Pi_a$, 
$k \in \Pi_b \cap \Pi_c$ following the boundary conditions described above. 
Given such a topological sector indexed by $\vec{n}$, we expect a discrete 
number of holomorphic discs to exist, and we have indicated such multiplicity 
by $d_{\vec{n}}$. The main contribution comes from the exponentiation of 
$A_{ijk}(\vec{n}) = \int_{\vec{n}} \om$, which is the target-area of 
such 'triangular' surface, whereas $\phi_{ijk}(\vec{n})$ is the phase the 
string endpoints pick up when going around the disc boundary $\partial D$ 
(see next section). As in (\ref{super}), the sign depends on the discs 
wrapping holomorphic or antiholomorphic maps. Finally, $h_{qu}$ stands for 
the contribution coming from quantum corrections, i.e., fluctuations 
around the minimal area semiclassical solution. Just as in the closed 
string case \cite{orbif}, we expect such contributions to factorize from 
the infinite semiclassical sum.

At this point one may wonder what is the detailed mechanism by which the 
chiral fermions get their mass. That is, one may want to understand what is 
the D-brane analog of Higgs mechanism in this intersecting brane picture. 
The right answer seems to be {\it brane recombination},
studied from a geometrical viewpoint by Joyce \cite{joyce2}, and later in 
terms of D-brane physics in \cite{km,bbh,angel02'}. The connection of such 
phenomenon to the SM Higgs mechanism was addressed in \cite{cim2}. Here we 
will briefly sketch this line of thought from a general viewpoint. Consider 
two D6-branes wrapping two sL's $\Pi_\a$ and $\Pi_\b$ on a ${\bf CY_3}$ 
$\cam$, and further assume that they have the same phase, i.e., both are 
calibrated by the same real volume form $\Om_\th$ and thus preserve 
(at least) a common $\cn = 1$ in $D = 4$.
From the geometrical viewpoint, 
they lie in a marginal stability wall of $\cam$. This implies that we can 
marginally deform our configuration by 'smoothing out' the intersections 
$\Pi_a \cap \Pi_\b$, combining the previous two sL's into a third one 
$\Pi_\g$. This family of deformations will all be calibrated by the same 
real volume form $\Om_\th$, so that the total volume or tension of the 
system will be invariant. From the field theory point of view, this 
deformation translates into giving non-vanishing v.e.v.'s to the massless 
scalar fields at the intersections.

\EPSFIGURE{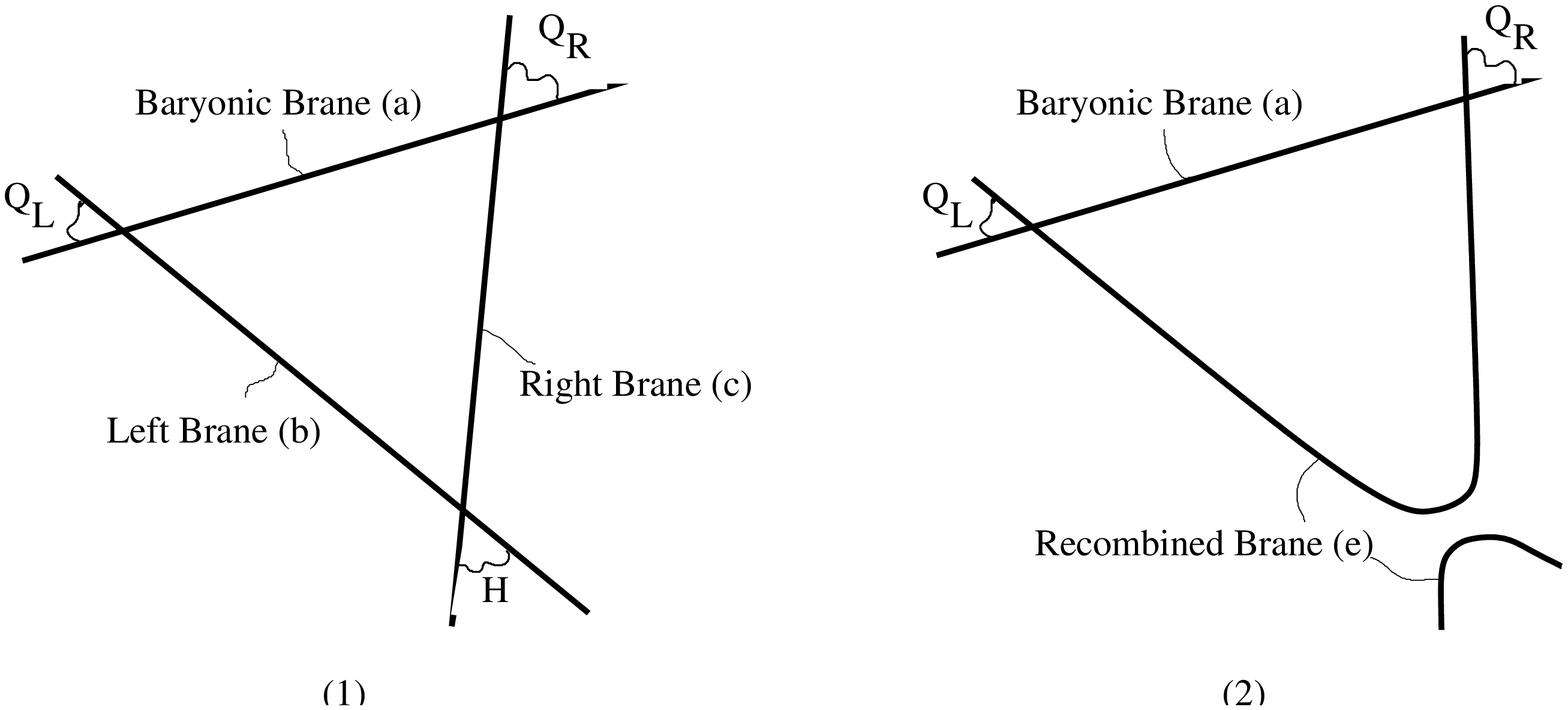, width=6in}
{\label{recombination}
Picture of the recombination. (1) Before the 
recombination, the worldsheet instantons connecting the $Q_L$, 
$Q_R$ and Higgs multiplets corresponds to holomorphic discs 
with their boundaries embedded in three different branes.
(2) After giving a v.e.v. to $H$, stacks $b$ and 
$c$ have recombined into a third one $e$.
If the recombination is soft enough, the number of 
chiral fermions at the intersections will not vary. 
However, they will get mass terms by holomorphic discs 
that connect fermions of opposite chirality, 
having its boundary on stacks $a$ and $e$.} 

Let us then consider the recombination of our SM stacks $b$ and $c$ into a 
third one $e$. By the above discussion, this correspond 
to giving a v.e.v. to Higgs multiplets living on $\Pi_b \cap \Pi_c$, so 
we expect that this implies a mass term for our chiral fermions. Indeed, in 
intersecting brane worlds, the chirality condition that prevents fermions 
from getting a mass is encoded in the non-vanishing topological intersection 
number of two branes, such as 
$[\Pi_a] \cdot [\Pi_b]$, $[\Pi_c] \cdot [\Pi_a] \neq 0$ that give us the 
number of net chiral quarks. Notice, however, that upon brane recombination, 
we will have 
\beq
\left|[\Pi_a] \cdot [\Pi_e]\right| = \left|[\Pi_a] 
\cdot ([\Pi_b] + [\Pi_c])\right| \leq \left|[\Pi_a] 
\cdot [\Pi_b]\right| + \left|[\Pi_a] \cdot [\Pi_c]\right|,
\label{recom}
\eeq
which implies that the number of 'protected' chiral fermions decreases if 
$[\Pi_a] \cdot [\Pi_b]$ and $[\Pi_a] \cdot [\Pi_c]$ have opposite sign, 
that is, yield fermions of opposite chirality. In realistic models we 
usually have $[\Pi_a] \cdot [\Pi_e] = 0$, so we expect every quark 
to get a mass. 

Since we are in a supersymmetric situation, we are allowed to perform an 
arbitrary small deformation from the initial configuration where the branes 
were not recombined. Upon such 'soft recombination', the actual number of 
intersections will not change, i.e., 
$\#(\Pi_a \cap \Pi_e) = \#(\Pi_a \cap \Pi_b) + \#(\Pi_a \cap \Pi_c)$. 
This implies that left and right-handed quarks will still be localized 
at intersections of $a$ and $e$. They will get, however, a mass term from 
a worldsheet instanton connecting each pair of them, now involving only 
two different boundaries. This situation has been illustrated in figure 
\ref{recombination}.

Before closing this section, let us mention that the discussion of Yukawa 
couplings, involving three or more stacks of branes, is intimately related 
to the previous discussion involving one single D-brane. Indeed, given a 
supersymmetric configuration of three stacks of D6-branes, we could think 
of slightly smoothing out each single intersection between each pair of 
them, thus recovering one single D6-brane wrapping a special Lagrangian. 
Now, by our general considerations of the superpotential of one single 
brane, we know that such superpotential will only depend on closed string 
K\"ahler moduli, and that will have the general form (\ref{super}). We 
expect the same results to hold in the case of the superpotential involving 
Yukawa couplings before recombination. In the next section we will compute 
such trilinear couplings for the simple case of Lagrangian $T^n$ wrapping 
n-cycles on $T^{2n}$, and see that they indeed satisfy such conditions.

\section{The general form of Yukawa couplings in toroidal models}

In this section we derive the general expression for Yukawa couplings in 
toroidal and factorizable intersecting brane configurations. By this we 
mean that the compact manifold will be a factorizable flat torus 
$\cam = T^{2n} = \otimes_{r=1}^n T_r^2$, whereas D-branes will be wrapping 
Lagrangian factorizable $n$-cycles, that is, those that can be expressed as 
a product of $n$ 1-cycles $\Pi_\a = \otimes_{r=1}^n (n_\a^r,m_\a^r)$, one 
on each $T^2$. Such $n$-cycles have the topology of $T^n$ and, if we minimize 
their volume in its homology class, they are described by hyperplanes 
quotiented by a torus lattice. This implies, in particular, that the 
intersection number between two cycles is nothing but the (signed) number of 
intersections, that is, $\#(\Pi_\a \cap \Pi_\b) = |[\Pi_\a] \cdot [\Pi_\b]|$. 
 Such class of configurations are known in the literature as 
{\it branes at angles} \cite{bdl}. Although our discussion in the 
previous section seems to indicate that the interesting case to study is 
$n = 3$, realistic models may be constructed involving also $n = 1, 2$ 
\cite{afiru}. For completeness, we derive our results for arbitrary $n$.

\subsection{Computing Yukawas on a $T^2$}

The simplest case when computing a sum of worldsheet instantons comes, as 
usual, from D-branes wrapping 1-cycles in a $T^2$, that is, branes 
intersecting at {\it one} angle. Let us then consider three of such 
branes, given by
\beq
\begin{array}{ccc}
\quad [\Pi_a] = [(n_a, m_a)] & \raw & z_a = R \cdot (n_a + \tau m_a) 
\cdot x_a  \\
\quad [\Pi_b] = [(n_b, m_b)] & \raw & z_b = R \cdot (n_b + \tau m_b) 
\cdot x_b \\
\quad [\Pi_c] = [(n_c, m_c)] & \raw & z_c = R \cdot (n_c + \tau m_c) 
\cdot x_c
\end{array}
\label{cycles}
\eeq
where $(n_\a,m_\a) \in \inte^2$ denote the 1-cycle the brane $\a$ wraps 
on $T^2$. 
Since the manifold of minimal volume in this homology class is given by a 
straight line with the proper slope, we can associate a complex number $z_\a$ 
to each brane, which stands for a segment of such 1-cycle in the covering 
space $\cpx$. Here $\tau$ is the complex structure of the torus and 
$x_\a \in \real$ an arbitrary number. We fix the area of 
$T^2$ (the K\"ahler structure, if we ignore the possibility of a B-field) 
to be $A = R^2 \ \pim \tau$. The triangles that will contribute to a 
Yukawa coupling involving branes $a$, $b$ and $c$ will consist of those 
triangles whose sides lie on such branes, hence of the form $(z_a, z_b, z_c)$.
 To be an actual triangle, however, we must impose that it {\it closes}, 
that is
\beq
z_a + z_b + z_c = 0.
\label{closure}
\eeq
Since $n_\a$, $m_\a$ can only take integer values, (\ref{closure}) can be 
translated into a Diophantine equation, whose solution is
\beqa
\begin{array}{c}
x_a = \left(I_{bc}/d\right) \cdot x \\ 
x_b = \left(I_{ca}/d\right) \cdot x \\ 
x_c = \left(I_{ab}/d\right) \cdot x
\end{array}
& \ {\rm with} &
\begin{array}{c}
x = \left(x_0 + l\right) \\
x_0 \in \real, \ l \in \inte \\
d = g.c.d. \left( I_{ab}, I_{bc}, I_{ca} \right) 
\end{array}
\label{diophan}
\eeqa
where $I_{\a\b} = [\Pi_\a] \cdot [\Pi_\b] = n_\a m_\b - n_\b m_\a$ 
stands for the intersection number of branes $\a$ and $\b$, and 
$x_0$ is a continuous parameter which is fixed for a 
particular choice of intersection points and brane positions, being a 
particular solution of (\ref{closure}). If, for instance, we choose branes 
$a$, $b$ and $c$ to intersect all at the same point, then we must take 
$x_0 = 0$. The discrete parameter $l$ then arises from triangles 
connecting different points in the covering space $\cpx$ but the same points 
under the lattice identification that defines our $T^2$. In the language of
the section 2, $l$ indexes the elements of the relative homology class
$H_2^D (T^2, \Pi_a \cup \Pi_b \cup \Pi_c, ijk)$. We thus see that a
 given Yukawa coupling gets contributions from an infinite (discrete) number 
of triangles indexed by $l$. 

Let us describe the specific values that $x_0$ can take. First notice that 
each pair of branes will intersect several times, each of them in a different
point of $T^2$. Namely, we can index such intersection points by
\beq
\begin{array}{cc}
i = 0,1, \dots, |I_{ab}| -1, & i \in \Pi_a \cap \Pi_b \\
j = 0,1, \dots, |I_{ca}| -1, & j \in \Pi_c \cap \Pi_a \\
k = 0,1, \dots, |I_{bc}| -1, & k \in \Pi_b \cap \Pi_c
\label{indices}
\end{array}
\eeq
In general, $x_0$ must depend on the particular triplet $(i,j,k)$ of 
intersection points and on the relative positions of the branes. For 
simplicity, let us take the triplet of intersections $(0,0,0)$ to correspond 
to a triangle of zero area. That is, we are supposing that the three branes 
intersect at a single point, which we will choose as the origin of the 
covering space (see figure \ref{tri}). Then it can be shown that, given the
 appropriate indexing of the intersection points, there is a simple expression
 for $x_0$ given by
\beq
x_0 (i,j,k) = \frac{i}{I_{ab}} + \frac{j}{I_{ca}} + \frac{k}{I_{bc}},
\label{ijkd=1}
\eeq 
where $i$, $j$ and $k$ are defined as in (\ref{indices})
\footnote{Notice that, since for a given triplet $(i,j,k)$ we must 
consider all the solutions $x_0 (i,j,k) + l$, $l \in \inte$, the index 
$i$ is actually defined mod $|I_{ab}|$, same for the others indices.}.
In this latter expression we are supposing that $d = 1$, that is, that 
$I_{ab}$, $I_{bc}$ and $I_{ca}$ are coprime integers. This guarantees that 
there exist a triangle connecting every triplet $(i,j,k)$, and also a simple
 expression for $x_0$. The case $d \neq 1$ will be treated below.
An illustrative example of the above formula is shown in figure \ref{tri}, 
where 
a triplet of 1-cycles intersecting at the origin have been depicted, both in 
a square torus and in its covering space, and the intersections have been 
indexed in the appropriate manner so that (\ref{ijkd=1}) holds. 

\EPSFIGURE{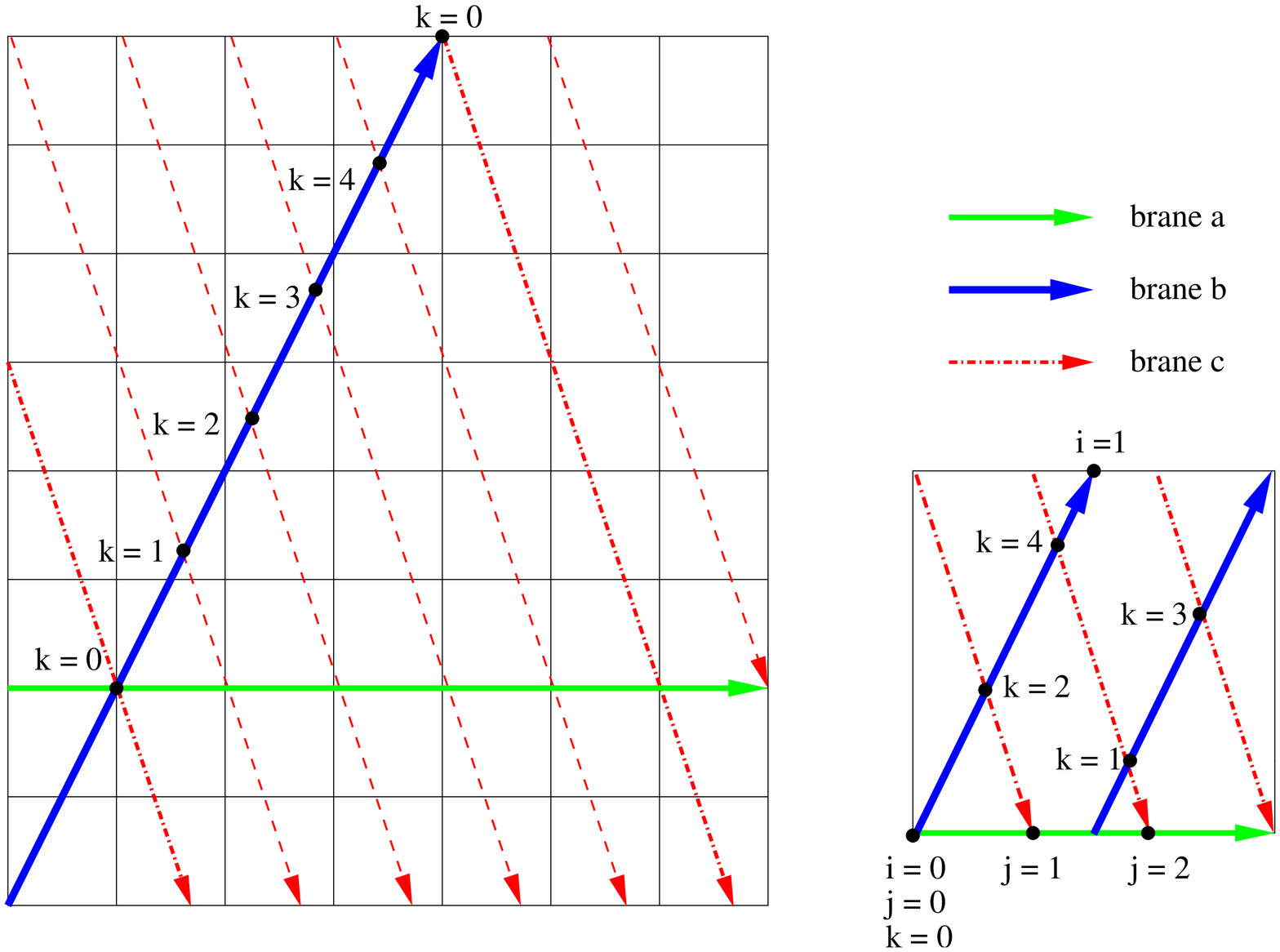, width=6in}
{\label{tri}
Relevant intersections and triangles 
for the three 1-cycles $(n_a,m_a) = (1,0)$, $(n_b,m_b) = (1,2)$ 
and  $(n_c,m_c) = (1,-3)$. In the left figure we have 
depicted these 1-cycles intersecting on the 
covering space $\cpx$. The dashed lines 
represent various images of the brane $c$ under 
torus translations. The indexing of the $bc$ intersections 
by the integer $k$ coincides with 
expressions (\ref{diophan}) and (\ref{ijkd=1}). In the right 
figure we have depicted a single fundamental region of the torus, 
and have indexed every 
intersection. Notice that we have chosen that all branes 
intersect at the origin, and the special choice of a square 
lattice for the complex structure. The results, however, are general.}

What if these branes do not intersect all at the origin? Let us consider 
shifting the positions of the three branes by the translations $\eps_a$, 
$\eps_b$ and $\eps_c$, 
where $\eps_\a$ is the transversal distance of the brane $\a$ from the origin 
measured in units of $A/ ||\Pi_\a||$, in clockwise sense from 
the direction defined by $\Pi_\a$. Then is easy to see that (\ref{ijkd=1})
 is transformed to
\beq
x_0 (i,j,k) = \frac{i}{I_{ab}} + \frac{j}{I_{ca}} + \frac{k}{I_{bc}}
+ \frac{I_{ab} \eps_c + I_{ca} \eps_{b} + I_{bc} \eps_{a}}
{I_{ab} I_{bc} I_{ca}}.
\label{ijkd=1eps}
\eeq 
Notice, however, that we can absorb these three parameters into only one, to 
be defined as $\tilde\eps = \frac{I_{ab} \eps_c + I_{ca} \eps_{b}
+ I_{bc} \eps_{a}}{I_{ab} I_{bc} I_{ca}}$. 
This was to be expected since, given the reparametrization invariance present 
in $T^2$, we can always choose branes $b$ and $c$ to 
intersect at the origin, and then the only freedom comes from shifting the 
brane $a$ away from this point.

Given this solution, now we can compute the areas of the triangles whose 
vertices lie on the triplet of intersections $(i,j,k)$ (we will say that this 
triangle 'connects' these three intersections), by using the well-known formula
\beq
A(z_a,z_b) = \oh \sqrt{|z_a|^2 \cdot |z_b|^2 - (\preal z_a \bar{z_b})^2 }.
\label{area}
\eeq 

Then we find that
\beqa
A_{ijk}(l) & = & \oh (2\pi)^2 A |I_{ab} I_{bc} I_{ca}| 
\ (x_0 (i,j,k) + l)^2 \nonumber \\
& = & \oh (2\pi)^2 A |I_{ab} I_{bc} I_{ca}| 
\left(\frac{i}{I_{ab}} + \frac{j}{I_{ca}} + 
\frac{k}{I_{bc}} + \tilde\eps + l \right)^2,
\label{area2}
\eeqa
where $A$ represents the K\"ahler structure of the torus,
and we have absorbed all the shift parameters into $\tilde\eps$. 
The area of such triangle may correspond to either an holomorphic 
or an antiholomorphic map from the disc. From (\ref{diophan}), we see
this depends on the sign of $I_{ab}I_{bc}I_{ca}$, so we must add 
a real phase $\sig_{abc} = {\rm sign } (I_{ab}I_{bc}I_{ca})$ 
to the full instanton contribution.

We can finally compute the corresponding Yukawa coupling for the 
three particles living at the intersections $(i,j,k)$:
\beq
Y_{ijk}  \sim  \sig_{abc} \sum_{l \in \inte}{\rm exp} 
\left( - {A_{ijk}(l) \over 2\pi \a^\prime} \right) 
\label{yukiT2}
\eeq

This last quantity can be naturally expressed in terms of a modular theta 
function, which in their real version are defined as
\beq
\vt \left[
\begin{array}{c}
\d \\ \phi
\end{array}
\right] (t) = \sum_{l \in \inte} q^{\oh (\d + l)^2} 
\ e^{2\pi i (\d + l)\phi}, \ \ q = e^{-2\pi t}.
\label{theta}
\eeq

Indeed, we find that (\ref{yukiT2}) can be expressed as such theta function 
with parameters
\beqa
\d & = & \frac{i}{I_{ab}} + \frac{j}{I_{ca}} + \frac{k}{I_{bc}}
+ \frac{I_{ab} \eps_c + I_{ca} \eps_{b} + I_{bc} \eps_{a}}
{I_{ab} I_{bc} I_{ca}}, \\
\phi & = & 0, \\
t & = & \frac{A}{\a^\prime} |I_{ab} I_{bc} I_{ca}|.
\label{paramT2}
\eeqa

\subsubsection{Adding a B-field and Wilson lines}

It is quite remarkable that we can express our Yukawa couplings in terms of a 
simple theta function. However, reached this point we could ask ourselves why 
it is such a specific theta function. That is, we are only considering the 
variable $t$ as a real number, instead of a more general parameter 
$\k \in \cpx$, and we are always setting $\phi = 0$. These two constraints 
imply that our theta functions are strictly real. From both the theoretical 
an phenomenological point of view, however, it would be interesting to have a 
Yukawa defined by a complex number. 

These two constraints come from the 
fact that we have considered very particular configurations of branes at 
angles. First of all we have not considered but tori where the B-field was 
turned off. This translates into a very special K\"ahler structure, where 
only the area plays an important r\^ole. In general, if we turn on a B-field, 
the string sweeping a two-dimensional surface will not only couple to the 
metric but also to this B-field. In a $T^2$, since the K\"ahler structure is
 the complex field
\beq
J = B + iA,
\label{kahler}
\eeq
we expect that, by including a B-field, our results (\ref{paramT2}) will 
remain almost unchanged, with the only change given by the substitution 
$A \raw (-i) J$. this amounts to changing our parameter $t$ to a complex one 
defined as 
\beq
\k = \frac{J}{\a^\prime} |I_{ab} I_{bc} I_{ca}|.
\label{cpxt}
\eeq

Our second generalization is including Wilson lines around the compact 
directions that the D-branes wrap. Indeed, when considering D-branes wrapping 
1-cycles on a $T^2$, we can consider the possibility of adding a Wilson line 
around this particular one-cycle. Since we do not want any gauge symmetry 
breaking, we will generally choose these Wilson lines to correspond to group 
elements on the centre of our gauge group, i.e., a phase 
\footnote{Notice that, although Wilson 
lines may produce a shift on the KK momenta 
living on the worldvolume of the brane, 
they never affect the mass of the particles 
living at the intersections, in the same 
manner that shifting the position of the branes
 does not affect them.}. 

Let us then consider a triangle formed by D-branes $a$, $b$ and $c$ each 
wrapped on one different 1-cycle of $T^2$ and with Wilson lines given by the 
phases $exp(2\pi i \th_a)$,  $exp(2\pi i \th_b)$ and $exp(2\pi i \th_c)$, 
respectively. The total phase that an open string sweeping such triangle picks 
up depends on the relative longitude of each segment, and is given by
\beq
e^{2\pi i x_a \th_a} \cdot e^{2\pi i x_b \th_b} \cdot e^{2\pi i x_c \th_c} 
= e^{2\pi i \left(I_{bc} \th_a + I_{ca} \th_b + I_{ab} \th_c \right) x}.
\label{wilson}
\eeq

Finally, we will consider both possibilities: having a B-field and some Wilson
 lines. In order to express our results we need to consider the complex theta 
function with characteristics, defined as 
\beq
\vt \left[
\begin{array}{c}
\d \\ \phi
\end{array}
\right] (\k) = \sum_{l \in \inte} 
e^{\pi i (\d + l)^2 \k} \ e^{2\pi i (\d + l) \phi }.
\label{thetacpx}
\eeq

Our results for the Yukawa couplings can then be expressed as such a function 
with parameters
\footnote{Notice that this implies that Yukawa couplings will be generically 
given by complex numbers, which is an important issue in order to achieve
a non-trivial CKM mixing phase in semirealistic models. }
\beqa
\d & = & \frac{i}{I_{ab}} + \frac{j}{I_{ca}} + \frac{k}{I_{bc}}
+ \frac{I_{ab} \eps_c + I_{ca} \eps_{b} + I_{bc} \eps_{a}}
{I_{ab} I_{bc} I_{ca}}, \\
\phi & = & I_{ab} \th_c + I_{ca} \th_b + I_{bc} \th_a,  \\
\k & = & \frac{J}{\a^\prime} |I_{ab} I_{bc} I_{ca}|.
\label{paramT2cpx}
\eeqa

\subsubsection{Orientifolding the torus}

From a phenomenological point of view, it is often  interesting to deal with
 a slight modification of the above toroidal model, which consist on
 performing an orientifold projection on the torus. Namely, we quotient the 
theory by $\OR$, where $\Om$ is the usual worldsheet orientation 
reversal and $\R: z \mapsto \bar z$ is a $\inte_2$ action on the torus. 
This introduces several new features, the most relevant for our discussion 
being

\begin{itemize}

\item There appears a new object: the O-plane, which lies on the 
horizontal axis described by $\{\pim z = 0 \}$ in the covering space $\cpx$.

\item In order to consider well-defined constructions, 
for each D-brane $a$ in our configuration we must 
include its image under $\OR$, denoted by $\Om\R a$ or $a^*$. 
These mirror branes will, generically, wrap a cycle $[\Pi_{a^*}]$ 
different from $
[\Pi_a]$, of course related by the action of $\R$ 
on the homology of the torus.

\end{itemize}

This last feature has a straightforward consequence, which is the 
proliferation of sectors as $ab$, $ab^*$, etc. Indeed, if we think of a 
configuration involving D-branes $a$, $b$ and $c$, we can no longer bother 
only about the triangle $abc$, but we must also consider $abc^*$, $ab^*c$ 
and $ab^*c^*$ triangles (the other possible combinations are mirror pairs of 
these four\footnote{We will not bother about triangles involving a brane and 
its mirror, as $abb^*$, for purely practical reasons. The results of this 
section, however, are easily extensible to these cases.}). Once specified the
 wrapping numbers of the triangle $abc$ all the others are also fixed. 
Since  our formulae for the Yukawas are not very sensitive to the actual 
wrapping number of the 1-cycles but only to the intersection numbers, we do 
not expect these to appear in the final expression. Notice, however, that if 
we specify the position of the brane $a$ the position of its mirror $a^*$ is 
also specified. Hence, shifts of branes should be related in the four 
triangles. This can be also deduced from the first item above. Since we have
 a rigid object lying in one definite 1-cycle, which is the O-plane, 
translation invariance is broken in the directions transverse to it, so we 
have to specify more parameters in a certain configuration. In this case of 
$T^2$ this means that if we consider that the three branes intersect at one 
point, we must specify the 'height' ($\pim z$) of such intersection. 

So our problem consists of, given the theta function parameters of the 
triangle $abc$, find those of the other three triangles. First notice that, 
if we actually consider the three branes $a$, $b$ and $c$ intersecting at one
 point, then the same will happen for the triangles $abc^*$, $ab^*c$ and 
$ab^*c^*$. Then by our previous results on triangles on a plain $T^2$ we see 
that the theta parameters will be given by
\beqa
\d_{abc} & = & \frac{i}{I_{ab}} + \frac{j}{I_{ca}} + \frac{k}{I_{bc}},\\
\k_{abc} & = & \frac{J}{\a^\prime} |I_{ab} I_{bc} I_{ca}|,
\label{paramoriabc}
\eeqa
for the $abc$ triangle and
\beqa
\d_{ab^*c} & = & \frac{i^*}{I_{ab^*}} + 
\frac{j}{I_{ca}} + \frac{k^*}{I_{b^*c}},\\
\k_{ab^*c} & = & \frac{J}{\a^\prime} |I_{ab^*} I_{b^*c} I_{ca}|,
\label{paramoriab*c}
\eeqa
for the $ab^*c$ triangle, etc. Notice that $i$ and $i^*$ are different indices
 which label, respectively, $ab$ and $ab^*$ intersections. 

A general configuration will not, however, contain every triplet of branes 
intersecting at one point, and will also contain non-zero Wilson lines. As 
mentioned, once specified the relative positions and Wilson lines of the 
triangle $abc$ all the other triangles are also specified. By simple 
inspection we can see that a general solution is given by the parameters  
\beqa
\d_{abc} & = & \frac{i}{I_{ab}} + \frac{j}{I_{ca}} + \frac{k}{I_{bc}}
+ \frac{I_{ab} \eps_c + I_{ca} \eps_{b} + I_{bc} \eps_{a}}
{I_{ab} I_{bc} I_{ca}}, \\
\phi_{abc} & = & I_{ab} \th_c + I_{ca} \th_b + I_{bc} \th_a,
\label{paramoriabceps}
\eeqa
for the triangle $abc$, and the parameters
\beqa
\d_{ab^*c} & = & \frac{i^*}{I_{ab^*}} + \frac{j}{I_{ca}} 
+ \frac{k^*}{I_{b^*c}}
+ \frac{I_{ab^*} \eps_c + I_{ca} \eps_{b^*} 
+ I_{b^*c} \eps_{a}}{I_{ab^*} I_{b^*c} I_{ca}}, 
\\
\phi_{ab^*c} & = & I_{ab^*} \th_c + I_{ca} \th_{b^*} + I_{b^*c} \th_a,
\label{paramoriab*ceps}
\eeqa
for the triangle $ab^*c$, and similarly for the other two triangles. Here we
 have defined
\beqa
\begin{array}{c}
\eps_{\a^*} = - \eps_\a \\
\th_{\a^*} = - \th_\a
\end{array}
 & & \a = a, b, c.
\label{mirrorparam}
\eeqa

\subsubsection{The non-coprime case}

Up to now, we have only consider a very particular class of Yukawa couplings:
 those that arise from intersecting D-branes wrapping 1-cycles on a $T^2$.
 Furthermore, we have also assumed 
the constraint $d = g.c.d. (I_{ab}, I_{bc}, I_{ca}) = 1$, 
that is, that the three intersection numbers are coprime. The 
non-coprime case is, however, the most interesting from the phenomenological 
point of view\footnote{This is no longer true when dealing with 
higher-dimensional cycles as, e.g., $n$-cycles wrapped on $T^{2n}$ for 
$n = 2,3$. In those cases, requiring that the brane configurations have only 
one Higgs particle imposes the coprime condition $d =1$ on each separate 
torus.}. In this section, we will try to address the non-coprime case. 
Although no explicit formula is given, we propose an ansatz that has been 
checked in plenty of models.

A particular feature of the configurations where $d > 1$ is that not every 
triplet of intersections $(i,j,k)$ is connected by a triangle. Indeed, from 
solution (\ref{diophan}) we see that a pair of intersections $(i,j)$ from 
$(ab,ca)$ will only couple to $|I_{bc}|/d$ different $bc$ intersections, same
 for the other pairs. Similarly, one definite intersection from $bc$ will 
couple to $|I_{ab}I_{ca}|/d^2$ $(i,j)$ pairs of $(ab,ca)$ intersections. This 
can be seen in figure \ref{tri2}, where a particular example of non-coprime 
configuration is shown.
\EPSFIGURE{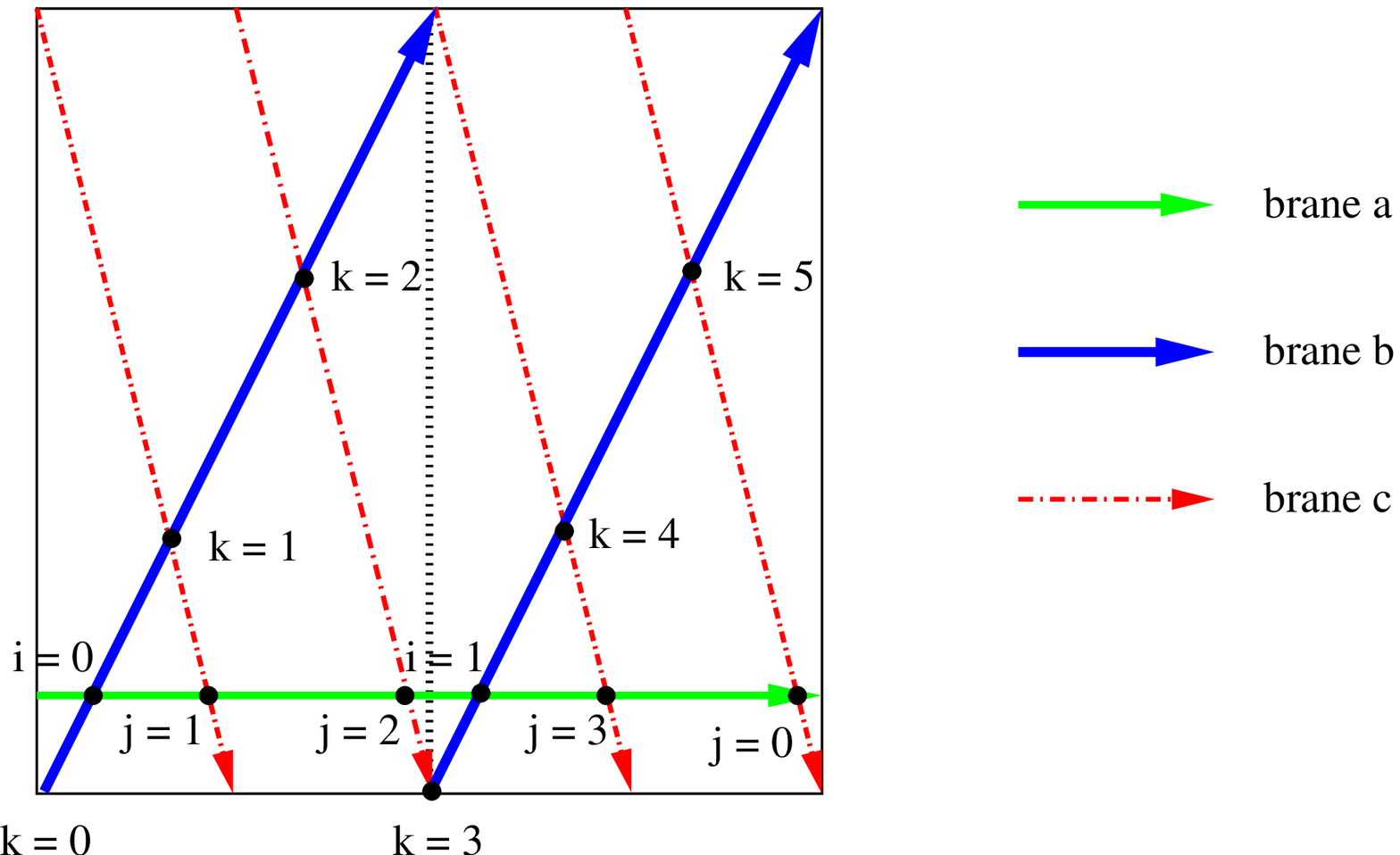, width=5in}
{\label{tri2}
Relevant intersections 
and triangles for the three 1-cycles $(n_a,m_a) = 
(1,0)$, $(n_b,m_b) = (1,2)$ and  $(n_c,m_c) = (1,-4)$. 
Notice that the fundamental region of the torus has two 
identical regions, exactly matching with $d = g.c.d. (I
_{ab}, I_{bc}, I_{ca}) = 2$. Also notice that a triangle 
exists connecting the vertices $(i,j,k)$ if and only if 
$i + j + k =$ even.}
In this same figure we can appreciate another feature of these configurations,
 which is that the fundamental region of the torus divides in $d$ identical 
copies. That is, the intersection pattern of any of these regions exactly 
matches the others. This is a direct consequence from the Diophantine 
solution (\ref{diophan}).

Let us now formulate the ansatz for this more general class of 
configurations. It consist of two points:
\begin{itemize}

\item A Yukawa coupling can be expressed as a complex 
theta function, whose parameters are
\beqa
\d & = & \frac{i}{I_{ab}} + \frac{j}{I_{ca}} + \frac{k}{I_{bc}}
+ \frac{d \cdot \left(I_{ab} \eps_c + I_{ca} \eps_{b} + 
I_{bc} \eps_{a}\right)}{I_{ab} I_{bc} I_{ca}} + \frac{s}{d},
\label{paramT2nc1}\\
\phi & = & \left( I_{ab} \th_c + I_{ca} \th_b + I_{bc} \th_a \right) / d,  
\label{paramT2nc2}\\
\kappa & = & \frac{J}{\a^\prime} {|I_{ab} I_{bc} I_{ca}| \over d^2} 
\label{paramT2nc3}
\eeqa
where $s \equiv s(i,j,k) \in \inte$ is a linear
function on the integers $i$, $j$ and $k$. 

\item A triplet of intersections $(i,j,k)$ 
is connected by a family of triangles, that is, 
has a non-zero Yukawa, if and only if
\beq
i + j + k \equiv 0 \ {\rm mod\ } d.
\label{condition}
\eeq

\end{itemize}

Notice that this ansatz correctly reduces to the previous solution in the 
coprime case, i.e., when $d = 1$. Indeed, in that case the actual value of 
$s$ becomes unimportant for the evaluation of the theta function and the 
condition (\ref{condition}) is trivially satisfied by any triplet $(i,j,k)$.

\subsection{Higher dimensional tori}

Having computed the sum of holomorphic instantons in the simple case of 
1-cycles in a $T^2$, let us now turn to the case of 
$T^{2n} = T^2 \times \dots \times T^2$. Since we are dealing with
 higher-dimensional geometry, surfaces are more difficult to visualize and 
computations less intuitive. We will see, however, that the final result is 
a straightforward generalization of the previous case. A $T^{2n}$ is a very 
particular case of ${\bf CY_n}$ manifold. Such manifolds are equipped with 
both a K\"ahler 2-form $\om$ and a volume $n$-form $\Om$ that satisfy
\beq
\frac{\om^n}{n!} = (-1)^{n(n-1)/2} \left( \frac i2\right)^n \Om \wedge \ov \Om.
\label{omegasn}
\eeq
In the particular case of a flat factorizable $T^{2n}$ we may take them to be
\beq
\om = \frac i2 \sum_{r=1}^n d z_r \wedge d\bar z_r \quad {\rm and} \quad
\Om = \preal \left( e^{i\th} d z_1 \wedge \dots \wedge d z_n \right).
\label{calit2n}
\eeq
As can easily be deduced from the discussion of section 2, these are not 
the only possible choices for $\Om$, but there are actually 2$^{n-1}$ 
independent complex $n$-forms satisfying (\ref{omegasn}), all suitable as 
calibrations. In particular, for suitable phases 
$\th_j$ ($j = 1, \dots, 2^{n-1}$) they all calibrate the so-called 
factorizable $n$-cycles, that is, the $n$-cycles that are Lagrangian $T^n$ 
and can be expressed as a product of $n$ 1-cycles 
$\Pi_\a = \otimes_{r=1}^n (n_\a^{(r)},m_\a^{(r)})$, one on each $T^2$. We 
will focus on configurations of branes on such factorizable cycles. Notice 
that these factorizable constructions, which yield branes intersecting at 
$n$ angles as in \cite{bdl}, are not the more general possibility. They are, 
however, particularly well-suited for extending our previous analysis of 
computation of Yukawas on a $T^2$. Indeed, the closure condition analogous 
to (\ref{closure}) can be decomposed into $n$ independent closure conditions, 
such as 
\beq
z_a^{(r)} + z_b^{(r)} + z_c^{(r)} = 0, \ \ r = 1, \dots, n,
\label{closure2n2}
\eeq
where $\a$ labels the corresponding $T^2$. Then we can apply our results from
 plain toroidal configurations to solve each of these $n$ Diophantine 
equations. The solution can then be expressed as three vectors 
$z_a, z_b, z_c \in \cpx^n$:
\beq
\begin{array}{ccc}
\quad [\Pi_a] = \bigotimes_{r = 1}^n [(n_a^{(r)}, m_a^{(r)})] 
& \quad \raw \quad & 
z_a =  \left( z_a^{(1)}, z_a^{(2)}, \dots, z_a^{(n)} \right), \\
\quad [\Pi_b] =  \bigotimes_{r = 1}^n [(n_b^{(r)}, m_b^{(r)})] 
& \quad \raw  \quad & 
z_b = \left( z_b^{(1)}, z_b^{(2)}, \dots, z_b^{(n)} \right), \\
\quad [\Pi_c] = \bigotimes_{r = 1}^n [(n_c^{(r)}, m_c^{(r)})] 
& \quad \raw \quad & 
z_c = \left( z_c^{(1)}, z_c^{(2)}, \dots, z_c^{(n)} \right).
\end{array}
\label{vectorscpx}
\eeq
Just as in (\ref{cycles}) and (\ref{diophan}), each entry is given by
\beqa
\begin{array}{c}
z_a^{(r)} = R^{(r)} \cdot \left(n_a^{(r)} + \tau^{(r)} m_a^{(r)}\right) 
I_{bc}^{(r)} x^{(r)} / d^{(r)} \\ 
z_b^{(r)} = R^{(r)} \cdot \left(n_b^{(r)} + \tau^{(r)} m_b^{(r)}\right)  
I_{ca}^{(r)} x^{(r)} / d^{(r)} \\ 
z_c^{(r)} = R^{(r)} \cdot \left(n_c^{(r)} + \tau^{(r)} m_c^{(r)}\right)  
I_{ab}^{(r)} x^{(r)} / d^{(r)}
\end{array}
& \quad {\rm with} &
\begin{array}{c}
x^{(r)} = \left(x_0^{(r)} + l^{(r)} \right) \\
x_0^{(r)} \in \real, \ l^{(r)} \in \inte \\
d^{(r)} = g.c.d. \left( I_{ab}^{(r)}, I_{bc}^{(r)}, I_{ca}^{(r)} \right) 
\end{array}
\label{diophan2n}
\eeqa
where $\tau^{(r)}$ denotes the complex structure of the corresponding 
two-torus, and the area of such is given by 
$A^{(r)} = (R^{(r)})^2 \cdot \pim \tau^{(r)}$. The intersection number of 
two cycles is simply computed as 
$I_{ab} = [\Pi_a] \cdot [\Pi_b] = \prod_{r = 1}^n I_{ab}^{(r)}$, where 
$I_{ab}^{(r)} = ( n_a^{(r)} m_b^{(r)} - n_b^{(r)} m_a^{(r)} )$ denotes 
the intersection number of cycles $a$ and $b$ on the $r^{th}$ $T^2$. Notice 
that now, each triplet of intersections $(i,j,k)$ is described by the 
multi-indices
\beq
\begin{array}{cc}
i = (i^{(1)}, i^{(2)}, \dots, i^{(n)}) \in \Pi_a \cap \Pi_b , 
& \quad i^{(r)} = 0, \dots, |I_{ab}^{(r)}| -1, \\
j = (j^{(1)}, j^{(2)}, \dots, j^{(n)}) \in \Pi_c \cap \Pi_a, 
& \quad j^{(r)} = 0, \dots, |I_{ca}^{(r)}| -1, \\
k = (k^{(1)}, k^{(2)}, \dots, k^{(n)}) \in \Pi_b \cap \Pi_c , 
& \quad k^{(r)} = 0, \dots, |I_{bc}^{(r)}| -1,
\end{array}
\label{multiindices}
\eeq
Correspondingly, each particular solution $x_0^{(r)}$ will depend on the 
triplet $(i^{(r)}, j^{(r)}, k^{(r)})$, and also on the corresponding 
shifting parameters. Namely,
\beq
x_0^{(r)} = \frac{i^{(r)}}{I_{ab}^{(r)}} 
+ \frac{j^{(r)}}{I_{ca}^{(r)}} 
+ \frac{k^{(r)}}{I_{bc}^{(r)}} 
+ \frac{d^{(r)} \cdot \left(I_{ab}^{(r)} \eps_c^{(r)} 
+ I_{ca}^{(r)} \eps_{b}^{(r)} + I_{bc}^{(r)} \eps_{a}^{(r)}\right)}
{I_{ab}^{(r)} I_{bc}^{(r)} I_{ca}^{(r)}}
+ \frac{s^{(r)}}{d^{(r)}}.
\label{ijkgeneral}
\eeq

Having parametrized the points of intersection in terms of the positions of 
the branes, it is now an easy matter to compute what is the area of the 
holomorphic surface that connects them. Recall that such a surface must have 
the topology of a disc embedded in $T^{2n}$, with its boundary embedded on 
the worldvolumes of $\Pi_a$, $\Pi_b$ and $\Pi_c$ (see figure \ref{yukis3}). 
Furthermore, in order to solve the equations of motion, it must be calibrated 
by $\om$ or, equivalently, parametrized by an (anti)holomorphic embedding 
into $T^{2n}$. We will discuss the existence and uniqueness of such surface 
in appendix A. For the time being, we only need to assume 
that it exist, since by properties of calibrations we know that its area is 
given by the direct evaluation of $\om$ on the relative homology class 
$H_2 (T^{2n}, \Pi_a \cup \Pi_b \cup \Pi_c, ijk) 
= \otimes_{r = 1}^n H_2 (T^2_r, \Pi_a^{(r)} \cup \Pi_b^{(r)} 
\cup \Pi_c^{(r)}, i^{(r)}j^{(r)}k^{(r)})$, indexed by the $n$ integer 
parameters $\{ l^{(r)} \}_{r = 1}^n$. Since $\om$ is essentially a sum of 
K\"ahler forms for each individual $T^2$, i.e., 
$\om = \sum_r \om_{T^2}^{(r)}$, this area is nothing but the sum of the 
areas of the triangles $(i^{(r)}j^{(r)}k^{(r)})$ defined on each $T^2$:
\beq
A(z_a,z_b) = \sum_r A(z_a^{(r)},z_b^{(r)}) = 
\oh (2\pi)^2 \sum_r  A^{(r)} 
\left|I_{ab}^{(r)} I_{bc}^{(r)} I_{ca}^{(r)}\right| 
\ \left(x_0^{(r)} + l^{(r)} \right)^2,
\label{areat2n}
\eeq
where we have used our previous computations (\ref{area}) and (\ref{area2}) 
relative to the case of $T^2$.

In order to compute the full instanton contribution, we must exponentiate 
such area as in (\ref{yukiT2}) and then sum over all the family of triangles. 
Notice that we must now sum over the whole of $n$ integer parameters 
$\{ l^{(r)} \}_{r = 1}^n$, one for each $T^2$. We thus find
\beqa
Y_{ijk}  & \sim &  \sig_{abc}
\sum_{\{l^{(r)}\} \in \inte^n}{\rm exp} 
\left( - {A_{ijk}(\{l^{(r)}\}) \over 2\pi \a^\prime} \right) =
\sig_{abc} \sum_{\{l^{(r)}\} \in \inte^n}{\rm exp} 
\left( - {\sum_\a A_{i^{(r)}j^{(r)}k^{(r)}}(l^{(r)}) 
\over 2\pi \a^\prime} \right) \nonumber \\
& = & \sig_{abc} \prod_r \sum_{l^{(r)} \in \inte} {\rm exp} 
\left( - {A_{i^{(r)}j^{(r)}k^{(r)}}(l^{(r)}) 
\over 2\pi \a^\prime} \right) = \sig_{abc}
\prod_r
\vt \left[
\begin{array}{c}
\d^{(r)} \\ 0
\end{array}
\right] (t^{(r)}),
\label{yukiT2n}
\eeqa
with $\d^{(r)} = x_0^{(r)}$ and 
$t^{(r)} =  A^{(r)}/\a^\prime |I_{ab}^{(r)} I_{bc}^{(r)} I_{ca}^{(r)}|$ 
as these {\it real} theta functions parameters. Here, 
$\sig_{abc} = \prod_r \sig^{(r)}_{abc} = 
\prod_r {\rm sign } (I_{ab}^{(r)} I_{bc}^{(r)} I_{ca}^{(r)}) = 
{\rm sign } (I_{ab}I_{bc}I_{ca})$. We thus see that for the case of higher 
dimensional tori, we obtain a straightforward generalization in terms of 
the $T^2$ case. Namely, the sum over worldsheet instantons is given by a product 
of theta functions. 

Given this result, is now an easy matter to generalize it to the case of 
non-zero $B$-field and Wilson lines. In order not to spoil the 
supersymmetric condition on D-branes wrapping sL's, we will add a 
non-vanishing $B$-field only in the dimensions transverse to them, that is, 
on the planes corresponding to each $T^2$. This complexifies the 
K\"ahler form to
\beq
 J^{(r)} = B_{(2r,2r+1)} + i A^{(r)}.
\label{cpxt2n}
\eeq
In the same manner, adding Wilson lines will contribute with a complex phase 
to the instanton amplitude. It can be easily seen that this phase will have 
the form
\beq
\prod_{r = 1}^n {\rm exp}\left( 2\pi i \left(I_{bc}^{(r)}
 \th_a^{(r)} + I_{ca}^{(r)} \th_b^{(r)} + I_{ab}^{(r)}
 \th_c^{(r)} \right) \cdot \left(x_0^{(r)} + l^{(r)} \right) \right),
\label{Wilson2n}
\eeq
where $\th_a^{(r)}$ correspond to a Wilson line of stack $a$ on the 
1-cycle wrapped on the $r^{th}$ $T^2$. These two sources of complex 
phases nicely fit into the definition of complex theta functions.

To sum up, we see that the Yukawa coupling for a triplet 
of intersections $(i,j,k)$ decomposed as in (\ref{multiindices}) 
will be given by
\beq
Y_{ijk} \sim \sig_{abc}
\prod_{r = 1}^n
\vt \left[
\begin{array}{c}
\d^{(r)} \\ \phi^{(r)}
\end{array}
\right] (\k^{(r)}),
\label{totalyuki}
\eeq
with parameters
\beqa
\d^{(r)} & = & \frac{i^{(r)}}{I_{ab}^{(r)}} 
+ \frac{j^{(r)}}{I_{ca}^{(r)}} 
+ \frac{k^{(r)}}{I_{bc}^{(r)}} 
+ \frac{d^{(r)} \cdot \left(I_{ab}^{(r)} \eps_c^{(r)} 
+ I_{ca}^{(r)} \eps_{b}^{(r)} + I_{bc}^{(r)} \eps_{a}^{(r)}\right)}
{I_{ab}^{(r)} I_{bc}^{(r)} I_{ca}^{(r)}}
+ \frac{s^{(r)}}{d^{(r)}},
\label{paramT2ncpx1}\\
\phi^{(r)} & = & 
\left(I_{ab}^{(r)} \th_c^{(r)} + 
I_{ca}^{(r)} \th_b^{(r)} + 
I_{bc}^{(r)} \th_a^{(r)}\right) / d^{(r)}, 
\label{paramT2ncpx2}\\
\k^{(r)} & = & 
\frac{J^{(r)}}{\a^\prime} 
\frac{|I_{ab}^{(r)} I_{bc}^{(r)} I_{ca}^{(r)}|}{(d^{(r)})^2}
\label{paramT2ncpx3}
\eeqa

\subsection{Physical interpretation}

Let us summarize our results. A Yukawa coupling between fields on the 
intersection of factorizable $n$-cycles $\Pi_a$, $\Pi_b$ and $\Pi_c$ on a 
factorizable $T^{2n}$ is given by 
\beq
Y_{ijk} = h_{qu} \sig_{abc}
\prod_{r = 1}^n
\vt \left[
\begin{array}{c}
\d^{(r)} \\ \phi^{(r)}
\end{array}
\right] (\k^{(r)}),
\label{totalyuki2}
\eeq
where $h_{qu}$ stands for the quantum contribution to the instanton 
amplitude. Such contributions arise from fluctuations of the worldsheet 
around the volume minimizing holomorphic surface. Given a triplet of 
factorizable cycles $abc$, the geometry of the several instantons are 
related by rescalings on the target space, so we expect these contributions 
to be the same for each instanton connecting a triplet 
$(i,j,k) \in (\Pi_a\cap \Pi_b,\Pi_c\cap \Pi_a, \Pi_b\cap \Pi_c)$, in close 
analogy with its closed string analogue \cite{orbif}. Moreover, such 
quantum contributions are expected to cancel the divergences that arise for 
small volumes of the compact manifold. Indeed, notice that by using the 
well-known property of the theta-functions
\beq
\vt \left[
\begin{array}{c}
\d \\ \phi
\end{array}
\right] (\k)
=
(-i\k)^{-1/2}
e^{2\pi i \d \phi}
\vt \left[
\begin{array}{c}
\phi \\ -\d
\end{array}
\right] (-1/\k),
\label{small}
\eeq
and taking $\k^{(r)} = \frac{i A^{(r)}}{\a^\prime} 
|I_{ab}^{(r)} I_{bc}^{(r)} I_{ca}^{(r)}|$, we see that $Y_{ijk}$ 
diverges as $( {\rm Vol} (T^{2n})/\a^\prime)^{-1/2}$.

Another salient feature of (\ref{totalyuki2}) involves its dependence in 
closed and open string moduli of the D-brane configuration. Notice that the 
only dependence of the Yukawa couplings on the closed string moduli enters 
through $J^{(r)}$, the K\"ahler structure of our compactification. Yukawa 
couplings are thus independent of the complex structure, which was to be 
expected from the general considerations of the previous section. On the 
other hand, the open string moduli are contained in the theta-function 
parameters $(\d^{(r)}, \phi^{(r)})$. If we define our complex 
moduli field as in the second ref. in \cite{kklm}, we find
\beq
\Phi_a^{(r)} = J \eps_a^{(r)} + \th_a^{(r)}
\label{omoduli}
\eeq
for the modulus field of D-brane wrapping $\Pi_a$, on the $r^{th}$ 
two-torus. By considering the K\"ahler moduli as external parameters, 
we recover Yukawa couplings which resemble those derived from a 
superpotential of the form (\ref{super}). Notice that not all the
 moduli are relevant for the value of the Yukawa couplings,  $2n$ of them 
decouple from the superpotential, as they can be absorbed by translation 
invariance in $T^{2n}$. The instanton generated superpotential will thus 
depend on $(K-2)n$ open moduli, where $K$ is the number of stacks of our 
configuration. In the orientifold case, only $n$ of such moduli decouple, 
so we have $(K-1)n$ such moduli. 


As a final remark, notice that our formula (\ref{totalyuki2}) has been 
obtained for the special case of a diagonal K\"ahler form $\om$. In the 
general case we would have
\beq
\om = \frac{i}{2} \sum_{r, s} a_{rs} \ dz_r \wedge d\bar z_s,
\label{gomega}
\eeq
so by evaluating $\om$ in the relative homology class we would expect an 
instanton contribution of the form
\beqa
Y_{ijk} & \sim & \sig_{abc} \
\vt \left[
\begin{array}{c}
\vec{\d} \\ \vec{\phi}
\end{array}
\right] ( A)  \nonumber \\
& = & \sig_{abc} \ \sum_{\vec{m} \in \inte^n} 
e^{i\pi (\vec{m} + \vec{\d}) \cdot A \cdot (\vec{m} + \vec{\d})}
e^{2\pi i (\vec{m} + \vec{\d}) \cdot  \vec{\phi}}
\label{multith}
\eeqa
where $A$ is an $n \times n$ matrix related to (\ref{gomega}) and
the intersection numbers of the $n$-cycles, and 
$\vec{\d},\ \vec{\phi}\  \in \real^n$ have entries defined by 
(\ref{paramT2ncpx1}, \ref{paramT2ncpx2}, \ref{paramT2ncpx3}). 
We thus see that the most general form of Yukawa 
couplings in intersecting brane worlds involves multi-theta functions, 
again paralleling the closed string case.

\section{An MSSM-like example}

\subsection{The model}

Let us illustrate the above general discussion with one specific example. 
In order to connect with Standard Model physics as much as possible, we will 
choose an intersecting brane model with a semi-realistic chiral spectrum, 
namely, that of the MSSM. As has been pointed out in \cite{imr}, it seems 
impossible to get an intersecting D6-brane model with minimal Standard 
Model-like chiral spectrum from plain toroidal or orbifold compactifications 
of type IIA string theory. One is thus led to perform an extra orientifold 
twist $\OR$ on the theory, $\Om$ being the usual worldsheet parity reversal 
and $\R$ a geometric (antiholomorphic) involution of the compact space 
$\cam$ \cite{CYberlin}. The set of fixed points of $\R$ will lead to the 
locus of an O6-plane \footnote{When dealing with orbifold constructions, 
several O-planes may appear. More precisely, each fixed point locus of 
$\R\iota$ with $\iota$ an element of the orbifold group satisfying 
$\iota^2 = 1$ will lead to an O-plane \cite{germans}.}. 

In addition, $\R: \cam \raw \cam$ will induce an action on the homology of 
$\cam$, more precisely on $H_3(\cam, \inte)$, where our D6-branes wrap.
\beq
\begin{array}{rc}
\R : & H_3(\cam, \inte) \raw H_3(\cam, \inte) \\
& [\Pi_\a] \mapsto [\Pi_{\a^*}]
\end{array}
\label{Rhomol} 
\eeq
Thus, as stated before, an orientifold configuration must consist of pairs 
($\Pi_\a$, $\Pi_{\a^*}$). If $\R(\Pi_\a) \neq \Pi_\a$, then a stack of $N_\a$ 
D6-branes on $\Pi_\a$ will yield a $U(N_\a)$ gauge group, identified with 
that on $\Pi_{\a^*}$ by the action of $\OR$ (i.e., complex conjugation). If, 
on the contrary, $\R(\Pi_\a) = \Pi_\a$, the gauge group will be real 
($SO(2N_\a)$) or pseudoreal ($USp(2N_\a)$).

We will use this simple fact when constructing our MSSM-like example. Indeed, 
notice that $USp(2) \cong SU(2)$, so in an orientifold setup $SU(2)_L$ weak 
interactions could arise from a stack of two branes fixed under $\R$. We 
will suppose that this is the case, which consists on a slight variation 
from the SM brane content of \cite{imr}. The new brane content is presented 
in table \ref{SMbranes} (see also figure \ref{sm}).
\TABLE{\renewcommand{\arraystretch}{1.5}
\begin{tabular}{|c|c|c|c|} 
\hline 
Label & Multiplicity & Gauge Group & Name \\ 
\hline 
\hline 
stack $a$ & $N_a = 3$ & $SU(3) \times U(1)_a$ & Baryonic brane\\ 
\hline 
stack $b$ & $N_b = 1$ & $SU(2)$ & Left brane\\ 
\hline 
stack $c$ & $N_c = 1$ & $U(1)_c$ & Right brane\\ 
\hline 
stack $d$ & $N_d = 1$ & $U(1)_d$ & Leptonic brane \\ 
\hline 
\end{tabular} 
\label{SMbranes}
\caption{\small Brane content yielding an MSSM-like spectrum. 
Only one representative of each brane is presented, not including 
the mirror branes $a^*, b^*, c^*, d^*$. Although $N_b = 1$, the 
mirror brane of $b$ lies on top of it, so it actually be considered
as a stack of two branes which, under $\Om$ projection, yield a 
$USp(2) = SU(2)$ gauge group.}}

Given this brane content, we can construct an intersecting brane model with 
the chiral content of the Standard Model (plus right-handed neutrinos) just 
by considering the following intersection numbers
 \beq 
\begin{array}{lcl} 
I_{ab}\ = \ 3, \\ 
I_{ac}\ = \ -3, & & I_{ac^*}\ =\ -3, \\ 
I_{db}\ = \ 3, \\ 
I_{dc}\ = \ -3, & & I_{dc^*}\ =\ -3,
\end{array} 
\label{intersec} 
\eeq 
all the other intersection numbers being zero (we have not included those 
involving $b = b$*). This chiral spectrum and the relevant non-abelian and 
$U(1)$ quantum numbers have been represented in table \ref{mssm}, together 
with their identification with SM matter fields.
\TABLE{\renewcommand{\arraystretch}{1.2}
\begin{tabular}{|c|c|c|c|c|c|c|}
\hline Intersection &
 SM Matter fields  & $SU(3) \ti SU(2)$  &  $Q_a$   & $Q_c $ & $Q_d$  & Y \\
\hline\hline (ab) & $Q_L$ &  $3(3,2)$ & 1   & 0 & 0 & 1/6 \\
\hline (ac) & $U_R$   &  $3( {\bar 3},1)$ & -1  &  1  & 0 & -2/3 \\
\hline (ac*) & $D_R$   &  $3( {\bar 3},1)$ &  -1  & -1    & 0 & 1/3
\\
\hline (db) & $L$    &  $3(1,2)$ &  0    & 0  & 1 & -1/2  \\
\hline (dc) & $N_R$   &  $3(1,1)$ &  0  & 1  &  -1  &  0  \\
\hline (dc*) & $E_R$   &  $3(1,1)$ &  0  & -1  & -1    & 1 \\
\hline 
\end{tabular}
\label{mssm}
\caption{\small Standard model spectrum and $U(1)$ charges.
The hypercharge generator is defined as 
$Q_Y = \frac 16 Q_a - \frac 12 Q_c - \frac 12 Q_d$.}}

Is easy to see that this spectrum is free of chiral anomalies, whereas it 
has an anomalous $U(1)$ given by $U(1)_a + U(1)_d$. Such anomaly will be 
canceled by a Green-Schwarz mechanism, the corresponding $U(1)$ gauge boson 
getting a Stueckelberg mass \cite{afiru} \footnote{The phenomenology related 
to such massive $U(1)$'s in low $\a'$ scenarios has been analyzed 
in \cite{uunos}.}. The two non-anomalous $U(1)$'s can be identified with 
$(B-L)$ and the 3$^{rd}$ component of right-handed weak isospin. This 
implies that the low energy gauge group is in principle 
$SU(3) \ti SU(2) \ti U(1)_{B-L} \ti U(1)_R$, giving the SM gauge group plus 
an extra $U(1)$. However, in orientifold models it may well happen that 
non-anomalous $U(1)$'s get also a mass by this same mechanism, the details
 of this depending on the specific homology cycles $[\Pi_\a]$, 
$\a = a, b, c, d$ \cite{imr}. This implies that in some specific 
constructions we could have only the SM gauge group. The Higgs system, 
which should arise from the $bc$ and $bc^*$ sector, gives no net chiral 
contribution and thus it does not appear at this abstract level of the 
construction, its associated spectrum depending on the particular 
realization of (\ref{intersec}) in terms of homology cycles
(see below).

Notice that the intersection numbers (\ref{intersec}) allow for the 
possibility $[\Pi_{c}] =  [\Pi_{c^*}]$. This would mean that, at some 
points on the moduli space of configurations $\Pi_{c} =  \Pi_{c^*}$ 
and the  stack $c$ gauge group could be enhanced as 
$U(1)_c \raw SU(2)_R$, just as for stack $b$. We would then recover a 
left-right symmetric model, continuously connected to the previous Standard 
Model-like configuration. By the same token, we could have 
$[\Pi_{a}] =  [\Pi_{d}]$, so when both stacks lied on top of each other we 
would get an enhancement $SU(3) \ti U(1)_{B-L} \raw SU(4)$. Considering both 
possibilities, one is naturally led to a intersecting brane configuration 
yielding a Pati-Salam spectrum, as has been drawn schematically in figure 
\ref{guayps}.

\EPSFIGURE{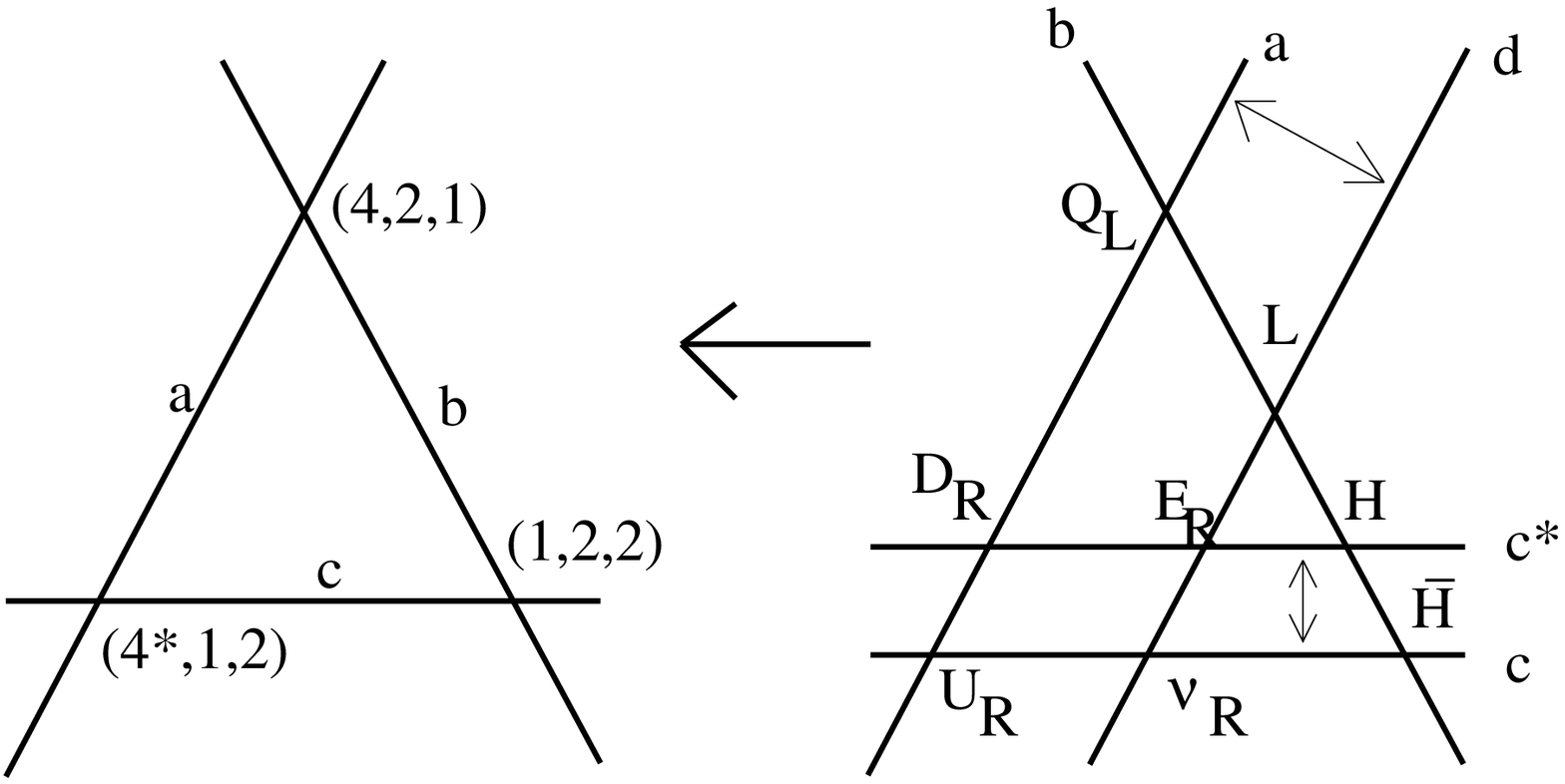, width=5in}
{\label{guayps}
Scheme of the model discussed in the text. Moving brane $c$ on top 
of its mirror $c$* one gets an enhanced $SU(2)_R$ symmetry.
If in addition brane $d$ is located on top of brane $a$ one gets
an enhanced $SU(4)$ Pati-Salam symmetry.}

Let us now give a specific realization of such abstract construction. For 
simplicity, we will consider a plain orientifold of type IIA compactified 
on a $T^6 = T^2 \ti T^2 \ti T^2$, with $\R: z_r \mapsto \bar z_r$ $r = 1,2,3$ 
being a simultaneous reflection on each complex plane. Our D6-branes will 
wrap factorizable cycles
\beq
[\Pi_\a] \equiv [(n_\a^1,m_\a^1)] \otimes [(n_\a^2,m_\a^2)]
\otimes [(n_\a^3,m_\a^3)],
\label{factorizable}
\eeq
and the action of $\R$ on such 3-cycles will be given by 
$\R : [(n_\a^{(r)}, m_\a^{(r)})] \mapsto [(n_\a^{(r)}, -m_\a^{(r)})]$ 
$r=1,2,3$, at least 
in square tori we will consider (for the action on tilted tori see 
\cite{bkl}). This compactification (again for square tori) possess 8 
different O6-planes, all of them wrapped on rigid 3-cycle in 
$[\Pi_{O6}] = \bigotimes_{r=1}^3 [(1,0)]^r$. This class of toroidal 
orientifold compactifications are related by T-duality with Type I D9 and 
D5-branes with magnetic fluxes \cite{bgkl}.

\TABLE{\renewcommand{\arraystretch}{1.5}
\begin{tabular}{|c||c|c|c|}
\hline
 $N_i$    &  $(n_\a^1,m_\a^1)$  &  $(n_\a^2,m_\a^2)$   & $(n_\a^3,m_\a^3)$ \\
\hline\hline $N_a=3$ & $(1,0)$  &  $(1/\rho , 3\rho )$ &
 $(1/\rho  ,  -3\rho )$  \\
\hline $N_b=1$ &   $(0, 1)$    &  $ (1,0)$  & 
$(0,-1)$   \\
\hline $N_c=1$ & $(0,1)$  & 
 $(0,-1)$  & $(1,0)$  \\
\hline $N_d=1$ &   $(1,0)$    &  $(1/\rho ,3\rho )$  &
$(1/\rho , -3\rho )$   \\
\hline \end{tabular}
\label{wnumbers}
\caption{\small D6-brane wrapping numbers giving rise to a
the chiral spectrum of the MSSM. Here $\rho = 1,1/3$.
The case $\rho = 1$ has been depicted in figure \ref{guay}.}}

A particular class of configurations satisfying (\ref{intersec}) in this 
specific setup is presented in table \ref{wnumbers}. A quick look at the 
wrapping numbers shows that this brane content by itself does not satisfy 
RR tadpole conditions $\sum_\a ([\Pi_\a] + [\Pi_{\a^*}]) = 32 [\Pi_{O6}]$. 
Although it does cancel all kind of chiral anomalies arising from the gauge 
groups in table \ref{SMbranes}, additional anomalies would appear in the 
worldvolume of D-brane probes as, e.g., D4-branes wrapping arbitrary 
supersymmetric 3-cycles \cite{angel00}. This construction should then be 
seen as a submodel embedded in a bigger one, where extra RR sources are 
included. These may either involve some hidden brane sector or NS-NS 
background fluxes, neither of these possibilities adding a {\it net} 
chiral matter content \cite{cim2}. As our main interest in this paper is 
giving a neat example where Yukawa couplings can be computed explicitly, 
we will not dwell on the details of such embedding.

Notice that this realization satisfies the constraints $[\Pi_a] = [\Pi_d]$ 
and $[\Pi_c] = [\Pi_{c^*}]$. Moreover, both $[\Pi_b]$ and $[\Pi_c]$ have a 
$USp(2N_\a)$ gauge group when being invariant under the orientifold action. 
This can easily be seen, since in the T-dual picture they correspond to 
Type I D5-branes, which by the arguments of \cite{gp} have symplectic gauge 
groups. As a result, this configuration of D-branes satisfies the conditions 
for becoming a Pati-Salam model in a subset of points of its open string 
moduli space (i.e., brane positions and Wilson lines). In addition, if we 
set the ratios of radii on the second and third tori to be equal (i.e., 
$R_2^{(2)}/R_1^{(2)} = R_2^{(3)}/R_1^{(3)} = \chi$) then one can check that
 the same $\cn = 1$ SUSY is preserved at each intersection \cite{csu,cim1}. 
Each chiral fermion in table \ref{mssm} will thus be accompanied by a scalar 
superpartner, yielding an MSSM-like spectrum.

Let us finally discuss the Higgs sector of this model. As mentioned before, 
stacks $b$ and $c$ correspond, in a T-dual picture, to two (dynamical) 
D5-branes wrapped on the second and third tori, respectively. Both D5's 
yield a $SU(2)$ gauge group when no Wilson lines are turned on their 
worldvolumes and, if they are on top of each other in the first torus, the 
massless spectrum in their intersection amounts to a $\cn = 2$ hypermultiplet 
in the representation $(2,2)$, invariant under CPT. This can also be seen as 
a $\cn = 1$ chiral multiplet. Turning back to the branes at angles picture 
we see that the intersection number $[\Pi_b]\cdot [\Pi_c] = 0$ because stacks 
$b$ and $c$ are parallel in the first torus, while they intersect only once 
in the remaining two tori. This single intersection will give us just one 
copy of the $(2,2)$ $\cn = 1$ chiral multiplet described above, whenever 
there exist the gauge enhancement to $SU(2) \ti SU(2)$. This will happen for 
stack $b$ whenever it is placed on top of any O6-plane on the second torus, 
and no Wilson line is turned on that direction. A similar story applies for 
stack $c$ in the third torus. Since we have no special interest in a gauge 
group $SU(2)_R$, we will consider arbitrary positions and Wilson lines for 
$c$ (see figure \ref{guay} for such a generic configuration). In that case, 
our $(2,2)$ chiral multiplet will split into $(2,-1)$ and $(2,+1)$ under 
$SU(2) \ti U(1)_c$, which can be identified with the MSSM Higgs particles 
$H_u$ and $H_d$, respectively. In addition, it exists a Coulomb branch 
between stacks $b$ and $c$ ($c$*), which corresponds to either geometrical 
separation in the first torus, either different 'Wilson line' phases along 
the 1-cycle wrapped on this $T^2$.\footnote{The complex phases associated 
to the 1-cycles of stacks $b$ and $c$ cannot be called Wilson lines in the 
strict sense, as they do not transform in the adjoint of $SU(2)$ but in 
the antisymmetric. 
This does not contradict section 2 general philosophy 
since such '1-cycles' are contractible in the orientifolded 
geometry.
} From the point of view of MSSM physics, these quantities can be 
interpreted as the real and imaginary part of a $\mu$-parameter, which is 
the only mass term for both Higgs doublets allowed by the symmetries of the 
model\footnote{Indeed, the associated term in the superpotential has been 
computed in the T-dual picture of Type I D5-branes in \cite{bl}, and shows 
the appropriate behaviour of a $\mu$-term.}.

After all these considerations, we see that the massless spectrum of table 
\ref{mssm} is that of the MSSM with a minimal Higgs set. Such a model was 
already presented in the third reference of \cite{reviews}, where some of 
its phenomenology as FI-terms were briefly studied. In the following, we 
will compute  the Yukawa couplings associated to such model which, as 
we will see, are particularly simple.

\subsection{Yukawa couplings}

Although we have given a explicit realization of (\ref{intersec}) by 
specifying the wrapping numbers of each stack of branes, the mere knowledge 
of the intersection numbers $I_{\a\b}^{(r)}$ on each $T^2_r$ would have been 
enough for computing the Yukawa couplings in this model. Indeed, the general 
formula (\ref{totalyuki}) only depends on these topological invariant 
numbers, plus some open string and closed string moduli. 

Let us first concentrate on the quark sector of the model, which involves 
the triplets of branes $abc$ and $abc$*. These correspond to the Up-like and 
Down-like quark Yukawas, respectively, as can be checked in table \ref{mssm}. 
Since stacks $b$ and $c$ are parallel in the first torus, the relative 
position and Wilson lines here do not affect the Yukawas (only the 
$\mu$-parameter). The Yukawa couplings will be given by the product of two 
theta functions, whose parameters depend on the second and third tori moduli. 
Let us take the option $\rho = 1$ in table \ref{wnumbers}. By applying 
formulae (\ref{paramT2ncpx1}) and (\ref{paramT2ncpx2}) we easily find these 
parameters for the triplet $abc$ to be
\beq
\begin{array}{rcl}
\d^{(2)} & = & 
- \frac {i^{(2)}}{3} - \left( \frac{\eps_a^{(2)}}{3} + \eps_c^{(2)} \right), \\
\d^{(3)} & = &
- \frac{j^{(3)}}{3} + \left(\frac{\eps_a^{(3)}}{3} - \eps_b^{(3)} \right) 
- \frac{\eps_c^{(3)}}{3},
\end{array}
\label{param1}
\eeq
\beq
\begin{array}{rcl}
\phi^{(2)} & = & - \left( \th_a^{(2)} + 3 \th_c^{(2)} \right), \\
\phi^{(3)} & = & \left( \th_a^{(3)} - 3 \th_b^{(3)} \right) - \th_c^{(3)},
\end{array}
\label{param2}
\eeq
where we have set $\eps_b^{(3)}$, $\th_b^{(3)}$ both to zero, in order to
 have the enhancement $U(1)_b \raw SU(2)_L$. In fact, their value must be 
frozen to either $0$ or $1/2$, so there are several possibilities, but all 
of them can be absorbed into redefinitions of the other continuous parameters.
Since the $abc$* triplet is related to $abc$ by orientifold reflection of 
one of its stacks, we can simply deduce its parameters by replacement 
$(\eps_c^{(3)},\th_c^{(3)}) \mapsto (-\eps_c^{(3)},-\th_c^{(3)})$, and 
$j^{(3)} \mapsto j^{(3)*}$ as the rules of section 3.1.2 teach us.
\TABLE{\renewcommand{\arraystretch}{1.5}
\begin{tabular}{|c||c||c|}
\hline
 & $abc$ & $abc$* \\
\hline
\hline
$\d^{(2)}$ 
& $\frac i3 + \eps^{(2)}$ 
& $\frac i3 + \eps^{(2)}$ \\
\hline
$\d^{(3)}$ & 
$\frac j3 + \eps^{(3)} + \tilde\eps^{(3)} $ 
& $\frac {j^*}{3} + \eps^{(3)} - \tilde\eps^{(3)} $  
\\
\hline
\hline
$\phi^{(2)}$ & 
$\th^{(2)}$ & $\th^{(2)}$ \\
\hline
$\phi^3$ & $\th^{(3)} + \tilde\th^{(3)}$ & $\th^{(3)} - \tilde\th^{(3)}$
\\
\hline
\end{tabular}
\label{moduli}
\caption{\small Parameters in the MSSM-like model of table 
\ref{wnumbers}, for the case $\rho = 1$. $i$ labels left-handed quarks, 
whereas $j$, $j$* label up and down-like right-handed quarks respectively.}}
Since these open string moduli $\eps$ and $\th$ appear in very definite 
combinations, we can express everything in terms of new variables. These 
can be interpreted as the linear combination of chiral fields living on the 
branes worldvolumes that couple to matter in the intersections through Yukawa 
couplings. The discrete indices $i$, $j$, $j$*, which label such matter at 
the intersections, have also been redefined for convenience. The final result
 is presented in table \ref{moduli}, and the geometrical meaning of these new
 variables is shown in figure \ref{guay}. Notice that, by field redefinitions,
 we can always take our open string moduli to range in $\eps \in [0,1/3)$ and 
$\th \in [0,1)$.

\EPSFIGURE{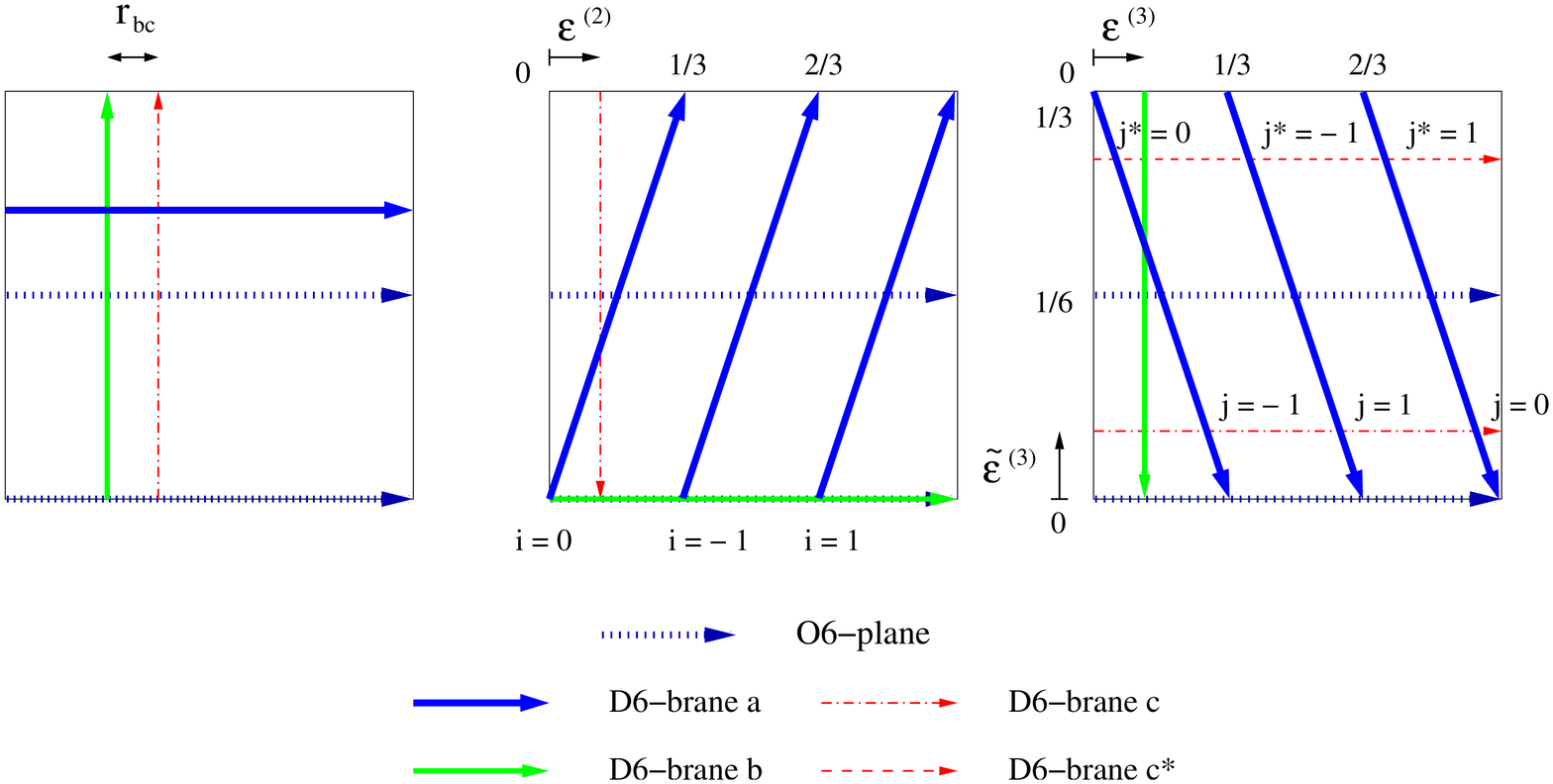, width=6.3in}
{\label{guay} Brane configuration corresponding to the
MSSM-like model described in the text, for the choice $\rho = 1$. 
For simplicity, we have not depicted the leptonic stack nor the mirror
$a$* stack.}

Considering the leptonic sector involves triplets $dbc$ and $dbc$*. Now, since
 the stack $d$ is similar to the $a$, the above discussion also apply to this 
case, and the only change that we have to make is considering new variables 
$(\eps^{(2)}_l, \eps^{(3)}_l; \th^{(2)}_l, \th^{(3)}_l)$ instead of 
$(\eps^{(2)}, \eps^{(3)}; \th^{(2)}, \th^{(3)})$. Notice that the difference 
of this two sets of variables parametrizes the breaking 
$SU(4) \raw SU(3) \ti U(1)_{B-L}$, whereas $(\eps_c^{(3)},\th_c^{(3)})$ 
parametrize $SU(2)_R \raw U(1)_c$ breaking. 

On the other hand, Yukawa couplings depend only on two closed string 
parameters, namely the complex K\"ahler structures on the second and third 
tori, through $\k^{(r)} = 3 J^{(r)}/\a' = 3 \chi (R^{(r)})^2/\a'$, 
$r = 2, 3$. Since the index $i$ is an index labeling left-handed quarks, 
whereas $j$, $j$* label up-like and down-like right-handed quarks, our 
Yukawa couplings will be of the form
\beq
Y_{ij}^U Q_L^i H_u U_R^j, \quad \quad Y_{ij*}^D Q_L^i H_d D_R^{j*},
\label{yukawas}
\eeq
with Yukawa matrices
\beq
\begin{array}{c}
Y_{ij}^U \sim 
\vt \left[
\begin{array}{c}
\frac i3 + \eps^{(2)} \\ \th^{(2)}
\end{array}
\right] \left({3 J^{(2)}  \over \a^\prime}\right) 
\times 
\vt \left[
\begin{array}{c}
\frac j3 + \eps^{(3)} + \tilde\eps^{(3)}\\ 
\th^{(3)} + \tilde\th^{(3)}
\end{array}
\right] \left({3 J^{(3)}  \over \a^\prime}\right), \\
Y_{ij*}^D \sim 
\vt \left[
\begin{array}{c}
\frac i3 + \eps^{(2)} \\ \th^{(2)}
\end{array}
\right] \left({3 J^{(2)}  \over \a^\prime}\right) 
\times 
\vt \left[
\begin{array}{c}
\frac {j*}{3} + \eps^{(3)} - \tilde\eps^{(3)}\\ 
\th^{(3)} - \tilde\th^{(3)}
\end{array}
\right] \left({3 J^{(3)}  \over \a^\prime}\right).
\end{array} 
\label{yukmatrices}
\eeq
We can apply analogous arguments for the case $\rho = 1/3$ in table 
\ref{wnumbers}. The final result is
\beq
Y_{ij}^U \sim 
\vt \left[
\begin{array}{c}
\frac j3 + \eps^{(2)} \\ \th^{(2)}
\end{array}
\right] \left({3 J^{(2)}  \over \a^\prime}\right)
\times 
\vt \left[
\begin{array}{c}
\frac i3 + \eps^{(3)} + \tilde\eps^{(3)}\\ 
\th^{(3)} + \tilde\th^{(3)}
\end{array}
\right] \left({3 J^{(3)}  \over \a^\prime}\right)
\label{yukmatrices2}
\eeq
for the up-like couplings, whereas the down-like ones are obtained form 
(\ref{yukmatrices2}) by the replacement 
$(\eps_c^{(3)},\th_c^{(3)}) \mapsto (-\eps_c^{(3)},-\th_c^{(3)})$.

The quark and lepton mass spectrum can be easily computed from these data. 
Indeed, let us consider the quark mass matrices proportional to 
(\ref{yukmatrices}), and define
\beq
\begin{array}{rcl}
a_i & \equiv & 
\vt \left[
\begin{array}{c}
\frac i3 + \eps^{(2)} \\ \th^{(2)}
\end{array}
\right] \left({3 J^{(2)}  \over \a^\prime}\right),
\\
b_j & \equiv &
\vt \left[
\begin{array}{c}
\frac j3 + \eps^{(3)} + \tilde\eps^{(3)}\\ 
\th^{(3)} + \tilde\th^{(3)}
\end{array}
\right] \left({3 J^{(3)}  \over \a^\prime}\right),
\\
\tilde b_{j*} & \equiv &
\vt \left[
\begin{array}{c}
\frac {j*}{3} + \eps^{(3)} - \tilde\eps^{(3)}\\ 
\th^{(3)} - \tilde\th^{(3)}
\end{array}
\right] \left({3 J^{(3)}  \over \a^\prime}\right).
\end{array}
\label{defin}
\eeq
Then, the Yukawa matrices can be expressed as
\beq
Y^U \sim A \cdot 
\left(
\begin{array}{ccc}
1 & 1 & 1 \\
1 & 1 & 1 \\
1 & 1 & 1
\end{array}
\right)
\cdot B,
\quad \quad
Y^D \sim A \cdot 
\left(
\begin{array}{ccc}
1 & 1 & 1 \\
1 & 1 & 1 \\
1 & 1 & 1
\end{array}
\right)
\cdot \tilde B.
\label{redef}
\eeq

\beq
\begin{array}{ccc}
A = \left(
\begin{array}{ccc}
a_1 &     &  \\
    & a_0 &  \\
    &     &  a_{-1}
\end{array}
\right) \quad
&
B = \left(
\begin{array}{ccc}
b_1 &     &  \\
    & b_0 &  \\
    &     &  b_{-1}
\end{array}
\right) \quad
&
\tilde B = \left(
\begin{array}{ccc}
\tilde b_1 &     &  \\
    & \tilde b_0 &  \\
    &     &  \tilde b_{-1}
\end{array}
\right).
\end{array}
\label{defin2}
\eeq

In order to compute the mass eigenstates, we can consider the hermitian, 
definite positive matrix $Y\cdot Y^\dag$ and diagonalize it. Let us take, 
for instance, $Y^U$. We find
\beq
Y^U \cdot (Y^U)^\dag \sim \Tr (B\cdot \bar B) \quad
A \cdot
\left(
\begin{array}{ccc}
1 & 1 & 1 \\
1 & 1 & 1 \\
1 & 1 & 1
\end{array}
\right)
\cdot \bar A,
\label{square}
\eeq
where bar denotes complex conjugation.
This matrix has one nonzero eigenvalue given by 
\beqa
\lam^U = \Tr (A\cdot \bar A) \ \Tr (B\cdot \bar B),
& \quad & 
| \lam^U \rangle = {A \over \sqrt{\Tr (A\cdot \bar A)}} \cdot 
\left(
\begin{array}{c}
1 \\ 1 \\ 1
\end{array}
\right),
\label{masseigenval}
\eeqa
and two zero eigenvalues whose eigenvectors span the subspace
\beq
\bar A^{-1} \cdot 
\left[{\rm Ker}
\left(
\begin{array}{ccc}
1 & 1 & 1 \\
1 & 1 & 1 \\
1 & 1 & 1
\end{array}
\right)
\right]
\bigcup
{\rm Ker} \bar A.
\label{zeroeigenval}
\eeq
Similar considerations can be applied to $Y^D$, and the
results only differ by the replacement $B \raw \tilde{B}$. This provides a 
natural mass scale between up-like and down-like quarks:
\beq
{m_U \over m_D} \sim 
\sqrt{\frac{\Tr (B\cdot \bar B)}{\Tr (\tilde{B}\cdot \bar{\tilde{B}})}}
\eeq
(we should also include $\langle H_u \rangle / \langle H_d \rangle$ in order 
to connect with actual quark masses). This ratio is equal to one whenever 
$\tilde \eps^{(3)} = - \tilde \eps^{(3)} \ {\rm mod} \ 1/3$ and 
$\tilde \th^{(3)} = - \tilde \th^{(3)} \ {\rm mod} \ 1$. These points in 
moduli space correspond precisely to the enhancement $U(1)_c \raw SU(2)_R$, 
where we would expect equal masses for up-like and down-like quarks. On the 
other hand, we find that the CKM matrix is the identity at every point in the 
moduli space. 

Thus we find that in this simple model only the third generation
of quarks and leptons are massive. This could  be considered 
as a promising starting point for a phenomenological
description of the SM fermion mass spectrum. One can conceive that
small perturbations of this simple brane setup can give
rise to smaller but non-vanishing masses for the rest of quarks 
and leptons as well as non-vanishing CKM mixing
\footnote{Note, for example, that the symmetry properties of
the Yukawa couplings leading to the presence of two massless modes 
disappear in the presence of a small
non-diagonal  component of the K\"ahler form $\omega$ as discussed at the
end of section 3.}. We postpone a detailed study of Yukawa couplings
in semirealistic intersecting D-brane models to future work.

The previous discussion parallels for the case $\rho = 1/3$. Indeed, we find 
the same mass spectrum of two massless and one massive eigenvalue for each 
Yukawa matrix. The only difference arises from the CKM matrix, which is not 
always the identity but only for the special values of 
$(\tilde \eps^{(3)}, \tilde \th^{(3)})$ where the symmetry enhancement to 
$SU(2)_R$ occurs.


\section{Extension to elliptic fibrations}

Although so far we have concentrated on computing Yukawa couplings 
in toroidal compactifications, it turns out that the same machinery
can be applied to certain D-brane models involving non-trivial 
Calabi-Yau geometries. Indeed, in \cite{local} a whole family of
intersecting D6-brane models wrapping 3-cycles of non-compact
${\bf CY_3}$'s was constructed. The simplest of such local Calabi-Yau 
geometries was based on elliptic and $\cpx$* fibrations over a 
complex plane, parametrized by $z$. 
In this setup, gauge theories arise from D6-branes 
wrapping compact special Lagrangian 3-cycles which, roughly speaking,
consist of real segments in the complex $z$-plane over which
two $S^1$ are fibered. One of such $S^1$ is a non-contractible cycle in 
$\cpx$*, while the other wraps a $(p,q)$ 1-cycle on the elliptic fiber.
The intersection of any such compact 3-cycles is localized on the base
point $z=0$, where the $\cpx$* fibre degenerates to $\cpx \ti \cpx$.
We refer the reader to \cite{local,otroslocal} for details 
on this construction. 

The important point for our discussion is that the geometry of any
number of intersecting D6-branes can be locally reduced to that of
intersecting 1-cycles on the elliptic fiber in $z=0$, that is, to that
of $(p_\a, q_\a)$ cycles on a $T^2$. Moreover, due to this local geometry,
any worldsheet instanton connecting a triplet of D6-branes will also 
be localized in the elliptic fiber at $z=0$. The computation of Yukawas
in this ${\bf CY_3}$ setup then mimics the one studied in section 3.1, 
where we considered worldsheet instantons on a $T^2$. 

As a result, we find that the structure of Yukawa couplings computed 
in section 3, which could be expressed in terms of (multi) theta 
functions, is in fact more general than the simple case of factorizable
cycles in a $T^{2n}$. In fact, it turns out to be even more general than
intersecting brane worlds setup. Indeed, as noticed in \cite{local}, 
this family of non-compact ${\bf CY_3}$ geometries is related, via
mirror symmetry, to Calabi-Yau threefold singularities given by complex
cones over del Pezzo surfaces. In turn, the intersecting D6-brane content
corresponds to D3-branes sitting on such singularities.

Let us illustrate these facts with a simple example already discussed 
in \cite{local}, section 2.5.1. The brane content consist of 
three stacks of $N$ branes each, wrapping the 1-cycles
\beq
\Pi_a = (2,-1), \quad \Pi_b = (-1,2), \quad  \Pi_c = (-1,-1),
\label{1cycles}
\eeq
with intersection numbers $I_{ab} = I_{bc} = I_{ca} = 3$. This
yields a simple $\cn=1$ spectrum with gauge group $U(N)^3$ and 
matter triplication in each bifundamental. We have 
depicted the D-brane content of (\ref{1cycles}) in figure \ref{local1}, 
restricting ourselves to the elliptic fiber on the base point $z=0$. 
Notice that the complex structure of such $T^2$ is fixed by the
$\inte_3$ symmetry that the whole configuration must preserve 
\cite{local}.
\EPSFIGURE{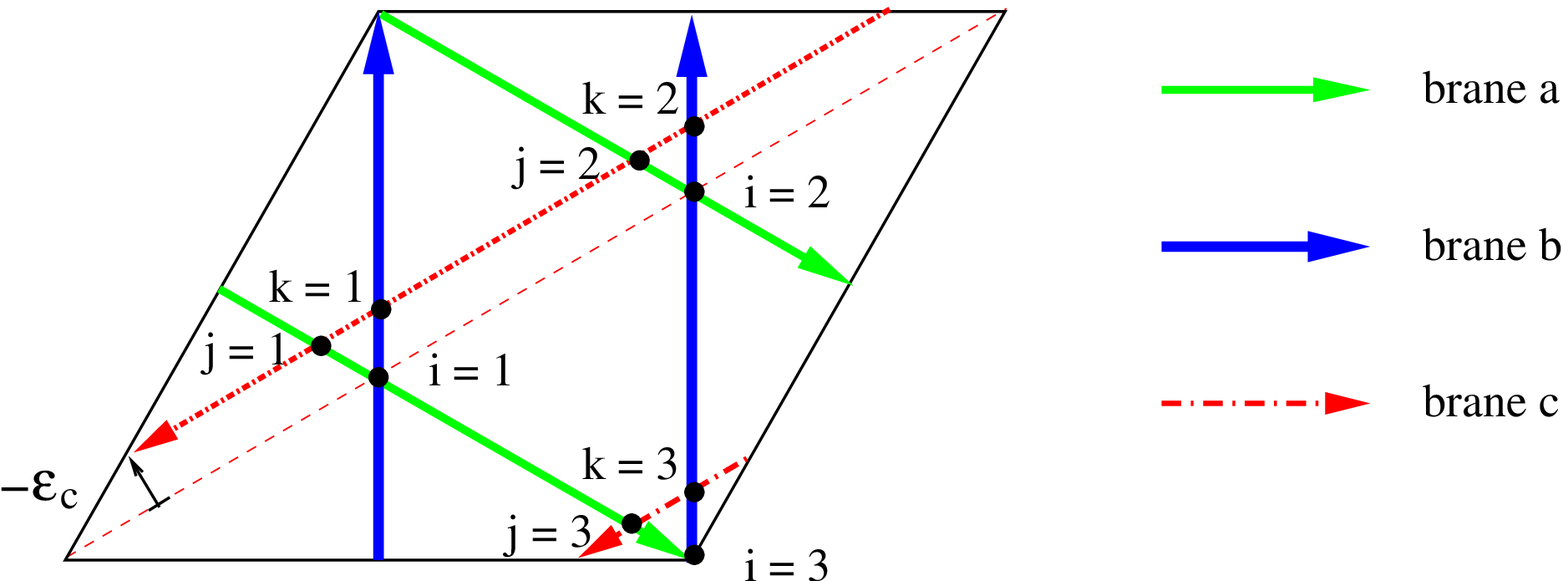, width=6in}
{\label{local1}
D-brane configuration in (\ref{1cycles}), restricted to
the elliptic fibre at $z=0$ in the base space.
Due to the quantum $\inte_3$ symmetry of this configuration, 
each stack of branes must be related to the others by a 
$e^{\pm 2\pi i/3}$ rotation. Hence, the complex structure
of the elliptic fibre is frozen.
Notice that a triangle between vertices
$i$, $j$, $k$ closes only if the condition 
$i + j + k \equiv 0$ mod $3$ is satisfied.}
Notice also that the intersection numbers are not coprime, 
so the Yukawa couplings between intersections $i$, $j$, $k$ will
be given by a theta function with characteristics 
(\ref{paramT2nc1}), (\ref{paramT2nc2}) and (\ref{paramT2nc3}),
with $d = 3$. Given the specific choice of numbering of
figure \ref{local1}, we can take the linear function
$s$ to be $s = -k - 2j$. Moreover, not all the triplet 
of intersections are connected by an instanton, 
but they have to satisfy the selection rule
\beq
i + j + k \equiv 0 \ {\rm mod} \ 3.
\label{selection}
\eeq

This give us the following form for the Yukawa 
couplings in the present model
\beq
Y_{ij1} \sim 
\left(
\begin{array}{ccc}
A & 0 & 0 \\
0 & 0 & B \\
0 & C & 0
\end{array}
\right), \quad
Y_{ij2} \sim 
\left(
\begin{array}{ccc}
0 & 0 & C \\
0 & A & 0 \\
B & 0 & 0
\end{array}
\right), \quad
Y_{ij3} \sim 
\left(
\begin{array}{ccc}
0 & B & 0 \\
C & 0 & 0 \\
0 & 0 & A
\end{array}
\right),
\label{yukilocal}
\eeq
with
\beq
A = 
\vt \left[
\begin{array}{c}
\eps/3 \\ 3\th
\end{array}
\right] (3 J/ \a'), \quad
B = 
\vt \left[
\begin{array}{c}
(\eps - 1)/3 \\ 3\th
\end{array}
\right] (3 J/ \a'), \quad
C = 
\vt \left[
\begin{array}{c}
(\eps + 1)/3 \\ 3\th
\end{array}
\right] (3 J/ \a'),
\label{thetas}
\eeq
and where we have defined the parameters 
$\eps = \eps_a + \eps_b + \eps_c,\ 
\th = \th_a + \th_b + \th_c \in [0,1)$.

A particularity of these elliptically fibered 3-cycles which
the D6-branes wrap is that, topologically, they are 3-spheres.
This means they are simply connected and, by \cite{McLean},
their moduli space is zero-dimensional. This means that the
D-brane position parameter $\eps_\a$ is fixed, and the same story 
holds for $\th_\a$.
Although frozen, we do not know the precise value of these 
quantities and, presumably, different values will correspond
to different physics. 

This simple model with matter triplication is in fact mirror
to the $\cpx^3/\inte_3$ orbifold singularity and, indeed,
the chiral matter content exactly reproduces the one obtained
from D3-branes at that singularity, in $N$ copies of the fundamental 
representation \cite{local}. The superpotential of such 
mirror configuration is given by
\beq
W = \sum_{\{abc\}} \eps^{ijk} [\Phi_{ab}^i \Phi_{bc}^j \Phi_{ca}^k],
\label{superp}
\eeq
where $\{abc\}$ means that we have to consider all the cyclic orderings.
This superpotential implies Yukawa couplings of the form 
$Y_{ijk} \sim \eps^{ijk}$. We see that we can reproduce such result
in terms of the general solution (\ref{yukilocal}), 
only if one of the entries $A$, $B$ or $C$ vanishes. Let us take
$C \equiv 0$, which can be obtained by fixing the theta-function 
parameters to be
\beq
\eps = \oh, \quad \quad \th = \frac{2m+1}{6}, \ m \in \inte.
\label{C=0}
\eeq
Is easy to see that this condition also implies that $|A| = |B|$.
More precisely,
\beq
A = Z \cdot e^{2\pi i (m + \oh)\frac{1}{6}},
\quad \quad
B = Z \cdot e^{-2\pi i (m + \oh)\frac{1}{6}}, 
\quad \quad
Z \in \cpx.
\label{conq}
\eeq
Now, if we perform the relabeling 
\beq
\begin{array}{cc}
i:  & 1 \lraw 3 \\
j:  & 1 \lraw 2 \\
k:  & 2 \lraw 3 
\end{array}
\label{relab}
\eeq
(which preserves the condition (\ref{selection}))
we are led to Yukawa couplings of the form
\beq
Y_{ij1} \sim 
Z \cdot \left(
\begin{array}{ccc}
 0 & 0 & 0 \\
 0 & 0 & \bar \om \\
 0 & \om & 0
\end{array}
\right), \quad
Y_{ij2} \sim 
Z \cdot \left(
\begin{array}{ccc}
0 & 0 & \om \\
0 & 0 & 0 \\
\bar \om & 0 & 0
\end{array}
\right), \quad
Y_{ij3} \sim 
Z \cdot \left(
\begin{array}{ccc}
0 & \bar \om & 0 \\
\om & 0 & 0 \\
0 & 0 & 0
\end{array}
\right),
\label{yukilocal2}
\eeq
where $\om = {\rm exp} (2\pi i (m + 1/2)\cdot 1/6)$. By taking the choice
$m = 1$, we obtain $\om = i$, so that $Y_{ijk} \sim \eps^{ijk}$, as 
was to be expected from (\ref{superp}). There are, however, two other
inequivalent choices of $\th$, given by $m = 0, 2$. Is easy to check
that these two values yield the superpotentials corresponding to the two
choices of $\cpx^3/(\inte_3 \times \inte_3 \times \inte_3)$  
orbifold singularity with discrete torsion, 
which have the same chiral spectrum as a plain $\cpx^3/\inte_3$ 
orbifold 
\footnote{It is, however, far from clear that these
configurations are actually mirror to orbifolds 
$\cpx^3/(\inte_3 \times \inte_3 \times \inte_3)$
with discrete torsion. Further checks involving, e.g., 
matching of moduli spaces should be performed to test this
possibility.}.
We present such final configuration in figure \ref{local2}.
\EPSFIGURE{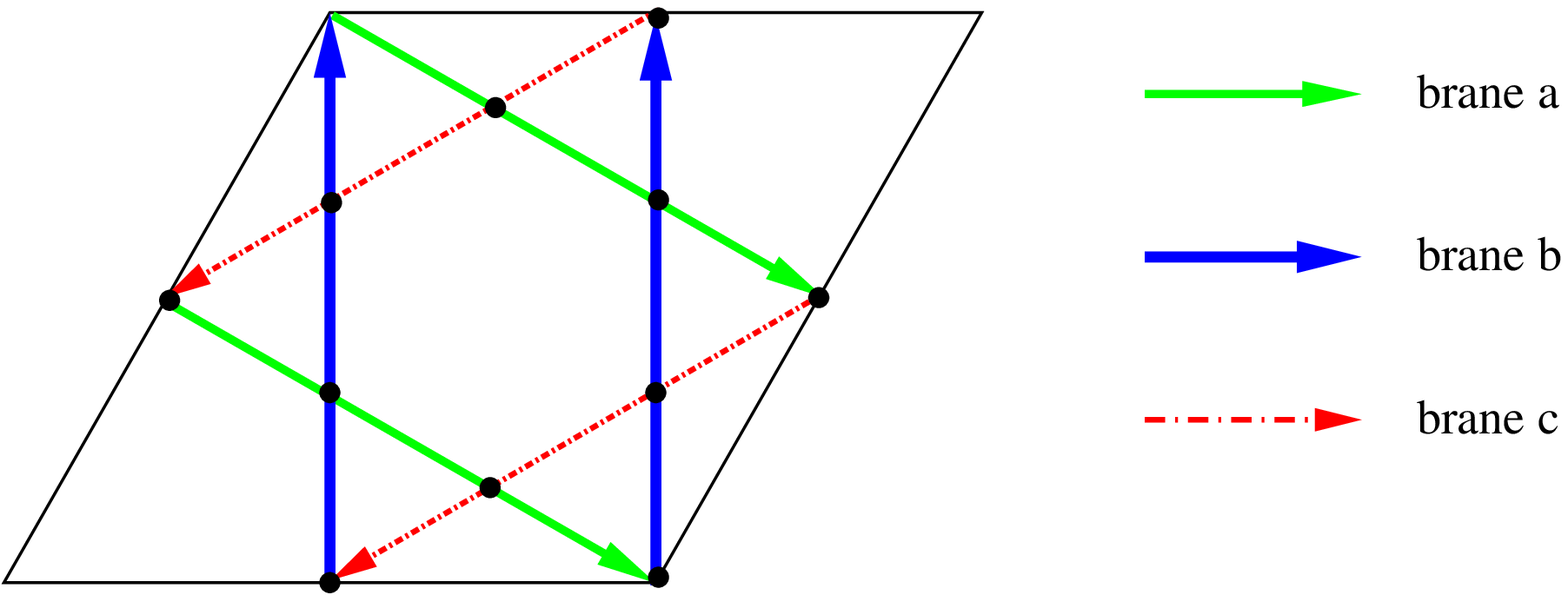, width=6in}
{\label{local2}
Final D-brane configuration, 
with the brane positions fixed by (\ref{C=0}).
Again we restrict to
the elliptic fibre at $z=0$ in the base space.}

\section{Yukawa versus Fukaya}

In the previous section we have shown how, combined with mirror symmetry,
the 
computation of worldsheet instantons between chiral fields in intersecting
D-brane models can yield a powerful tool to compute Yukawa couplings in 
more general setups as, e.g., D-branes at singularities. The purpose of
the present section is to note that computation of Yukawas and other 
disc worldsheet instantons is not only a tool, but lies at the very heart
of the definition of mirror symmetry. The precise context to look at is
Kontsevich's homological mirror symmetry conjecture \cite{Konty}, performed 
before the importance of D-branes was appreciated by the physics community.
This proposal relates two a priori very different constructions in two 
different $n$-fold Calabi-Yau manifolds $\cam$ and $\cw$, which are dual 
(or mirror) to each other. $\cam$ is to be seen as a $2n$-dimensional
symplectic manifold with vanishing first Chern class, while $\cw$ 
shall be regarded as an $n$-dimensional complex algebraic manifold.
From the physics viewpoint these are the so-called A-side and B-side 
of the mirror map, respectively. The structure of complex manifold
allows us to construct the derived category of coherent sheaves 
from $\cw$, which classifies the boundary conditions in the B-twisted
topological open string theory model on such manifold, hence the 
spectrum of BPS branes on the B-side \cite{cate}. On the other hand,
the symplectic structure on $\cam$ naturally leads to the construction
of Fukaya's category. Kontsevich's proposal amounts to the equivalence 
of both categories.
\footnote{Actually, it identifies a properly enlarged (derived) version
of Fukaya's category to a full subcategory of coherent sheaves. 
since we are only interested on the A-side of the story, 
we will not deal on these subtle points regarding the mirror map.}

Intersecting D6-brane configurations as described in section 2 lie on 
the A-side of this story. Hence, they should be described by Fukaya's
category. This seems indeed to be the case and, in the following, we 
will try to describe the physical meaning of Fukaya's mathematical 
construction from the point of view of intersecting D-brane models, 
paying special attention to the r\^ole of worldsheet instantons.

Roughly speaking, a category is given by a set of objects and morphisms
between them. The objects of the category we want to construct are
pairs of the form $\cl_\a = (\Pi_\a, \ce_\a)$, of 
special Lagrangian submanifolds $\Pi_\a$ of $\cam$, endowed with unitary 
local systems $\ce_\a$, or flat $U(N_\a)$ gauge bundles. 
This is precisely the set of objects
used as building blocks in our ${\bf CY_n}$ intersecting brane 
configurations, each object associated to a stack of $N_\a$ D-branes
\footnote{In the general construction, Fukaya considers a different 
(larger) set of objects in $\cam$, which are equivalence classes 
of Lagrangian submanifolds identified by Hamiltonian diffeomorphisms. 
The precise relation between the moduli space of these two 
set of objects is still an open problem, but in some simple cases 
as Lagrangian tori in $\cam = T^{2n}$ they can be shown to be 
equal \cite{theta}.}. 
The morphisms between a pair of objects are
generated by the points of intersection of the two sL submanifolds,
more precisely
\beq
Hom (\cl_\a, \cl_\b) = {\# (\Pi_\a \cap \Pi_\b)} \otimes Hom (\ce_\a, \ce_\b),
\label{morph}
\eeq
where $Hom (\ce_\a, \ce_\b)$ represents a morphism of gauge bundles. 
The morphisms between two objects, then, are naturally associated 
to the set of chiral fields living at the intersections of two D-branes, 
which transform under the bifundamental representation of the 
corresponding gauge groups.

Actually, (\ref{morph}) can be converted to a {\it graded} set of
morphisms. That is, each intersection $p$ can be attached with an 
index $\eta(p) \in \inte$, named Maslov index. The physical meaning of 
such index is related to how one must describe the spectrum of particles
at the intersections from the underlying CFT viewpoint, i.e., two
sets of fields arising from two elements of $\Pi_\a \cap \Pi_\b$ may
look locally similar, but related by some spectral flow that the 
global theory must take into account. In the case of Lagrangian tori,
i.e., factorizable D-branes, in $T^{2n}$ all the intersections 
have the same Maslov index, which can be defined as 
\cite{theta}
\beq
\eta (p)_{\a\b} = \sum_{r=1}^n [\th_{\a\b}^r],
\label{maslov}
\eeq
where $[\th_{\a\b}^r]$ is (the integer part of) the angle between
$\Pi_\a$ and $\Pi_\b$ on the $r^{th}$ two-torus, measured in the
counterclockwise sense and in units of $\pi$. A quick look at the
case $n = 3$ (D6-branes) reveals that $\eta (p)$ mod $2$ can be
associated to the chirality of the corresponding field at the 
intersection $p$. In fact, this grading was already present from the
very beginning since, actually, (\ref{morph}) is only valid whenever
$\Pi_\a$ and $\Pi_\b$ intersect transversally in $\cam$. In case their 
intersection is not a point, we should replace (\ref{morph}) 
by the cohomology complex $H^*(\Pi_\a \cap \Pi_\b, Hom (\ce_\a, \ce_\b))$, 
with its associate grading. We finally find that the morphisms of 
our construction are given by the graded set
\beq
Hom^i (\cl_\a, \cl_\b) \simeq \bigotimes_{p \in \Pi_\a \cap \Pi_\b} 
H^{i-\eta(p)} (\Pi_\a \cap \Pi_\b, Hom (\ce_\a, \ce_\b)).
\label{graded}
\eeq

In general, categories not only consist of objects and morphisms
between them, but also of compositions of morphisms. In \cite{Fuka}
Fukaya proved that, actually, an involved structure can be defined 
in the previous described category. This structure is based on the
properties of an $A^\infty$ algebra, which is a generalization of 
a differential graded algebra, and makes Fukaya's category into an
$A^\infty$ category. The definition of such category involves
\begin{itemize}

\item A class of objects

\item For any two objects $X$, $Y$ a \inte-graded Abelian group of
morphisms $Hom(X,Y)$

\item Composition maps
\beq
m_k : Hom(X_1,X_2) \otimes  Hom(X_2,X_3) \dots \otimes
 Hom(X_k,X_{k+1}) \raw  Hom(X_1,X_{k+1}),
\nonumber
\eeq
of degree $2-k$ for all $k \geq 1$, satisfying the $n^{th}$ order
associativity conditions

$\sum_{r,s} (-1)^\eps
m_{n-r+1} (a_1 \otimes \dots \otimes a_{s-1} \otimes 
m_r(a_s \otimes \dots \otimes a_{s+r-1})
\otimes a_{s+r} \otimes \dots \otimes a_n) = 0$

for all $n \geq 1$. Here 
$\eps = (s+1)k + s \left(n + \sum_{j=1}^{s-1} {\rm deg} (a_j)\right)$.

\end{itemize}

Notice that the associative condition for $n=1$ reduces to $m_1$ being
a degree one operator $m_1: Hom(X,Y) \raw Hom(X,Y)$ such that 
$m_1 \circ m_1 = 0$, hence it defines a coboundary operator 
which makes $Hom(X,Y)$ a cochain complex 
(e.g., just as the exterior derivative in the deRham complex). 
Furthermore, the second relation ($n=2$) means that 
$m_2: Hom(X,Y) \times Hom(Y,Z) \raw Hom(X,Z)$ is a cochain homomorphism,
which induces a product on cohomology groups 
(e.g., analogous to a wedge product in deRham cohomology). 

Up to now we have described the two first items of the previous definition,
and is in the third one where worldsheet instantons will play a r\^ole.
Indeed, the composition maps $m_k$ are constructed by considering 
holomorphic maps $\phi: D \raw \cam$, taking the boundary of 
$D$ to $k+1$ Lagrangian submanifolds $\cl_\a$ in $\cam$
and $k+1$ marked points $z_\a$ in the cyclic order of $\partial D$ 
to the intersection points $p_\a \in \Pi_\a \cap \Pi_{\a+1}$ 
(see figure \ref{yukis3} for the case $k = 2$). Two of these maps 
are regarded  as equivalent if related by a disc conformal automorphism
$Aut(D) \simeq PSL(2,\real)$. This is precisely the definition of a
euclidean worldsheet connecting $k+1$ fields at the intersection
of $k+1$ D-branes and satisfying the classical equations of motion. 
Let us fix $k+1$ intersections $p_\a \in \Pi_\a \cap \Pi_{\a+1}$, 
and the corresponding $k+1$ matrices 
$t_\a \in Hom(\ce_\a|_{p_\a}, \ce_{\a+1}|_{p_{\a+1}})$.
If we choose them to have the appropriate degree, the moduli space of 
holomorphic inequivalent maps connecting them will be discrete, 
so we can sum over all of such $\phi$ to obtain the quantity
\beq
C((p_1,t_1), \dots,(p_k,t_k), p_{k+1}) = \sum_{\phi} \pm 
\ e^{2\pi i \int_{\phi(D)} \om}
\ P e^{\oint_{\phi(\partial D)} \b},
\label{instanton}
\eeq
 where the $\pm$ sign stands for holomorphic and antiholomorphic maps,
respectively and $\om$ is the complexified K\"ahler form in $\cam$. 
Finally, $P$ stands for a path-ordered integration on the boundary of 
the disc, with $\b$ is being the connection of the flat bundles 
along the boundaries in $\cl_\a$'s, and the matrices $t_i$ inserted
at the boundary marked points \cite{poli}. The r.h.s. of 
(\ref{instanton}) is thus a homomorphism of $\ce_1$ to $\ce_{k+1}$,
and we can define the composition maps $m_k$ as
\beq
m_k ((p_1,t_1), \dots,(p_k,t_k)) = 
\sum_{p_{k+1} \in \Pi_1 \cap \Pi_{k+1}}
C((p_1,t_1), \dots,(p_k,t_k), p_{k+1}) \cdot p_{k+1}.
\label{comp}
\eeq

From these definitions, is easy to see that the computation
of the composition maps $m_k$ is equivalent to computing the
worldsheet instanton correction to the superpotential involving
$k+1$ chiral fields at the D-branes intersections. Indeed,
(\ref{instanton}) is nothing but a generalization of (\ref{yukabs})
for $k+1$ disc insertions. In the case $k=1$, we are evaluating
the instantons connecting two chiral fields at the intersections
of the same pair of D-branes $\cl_1$, $\cl_2$, just as in figure
\ref{recombination} (2). Notice that $m_1$ has degree $1$, so 
it can be regarded as a chirality-changing operator, i.e., a mass term. 
If we consider D-brane configurations consisting of Lagrangian $n$-tori 
on $T^{2n}$ its action is trivial, $m_1 \equiv 0$, but this will not
be the case in general. Computing the cohomology
\beq
{\{ {\rm Ker} 
\left(m_1: Hom^{i} (\cl_\a, \cl_\b)
\raw Hom^{i+1} (\cl_\a, \cl_\b) \right) \}
\over
\{ {\rm Im} 
\left(m_1: Hom^{i-1} (\cl_\a, \cl_\b)
\raw Hom^{i} (\cl_\a, \cl_\b) \right) \} },
\label{cohom}
\eeq
is equivalent to computing the massless fields in our D-brane 
configuration. On the other hand, the case $k=2$ coincides
with the general computation of Yukawa couplings between
chiral fields at sL's intersections described in section 2.3.
Indeed, the action of $m_2$ involves three objects of our 
configurations, hence three D-branes $(a, b, c)$, and three 
different morphisms between them, hence a triplet of intersections
$(i,j,k)$. For each such choice we can define an element of the form
(\ref{instanton}), which represents a Yukawa coupling. Indeed,
in the phenomenological setup of figure \ref{yukis3}, if we let
$i,j$ index left- and right-handed quarks, respectively, 
the matrix of maps $(M_2)_{ij} = m_2((p_i,t_i),(p_j,t_j))$
is equivalent to the Yukawa matrix of such chiral fields. 
In the same manner, the composition maps $m_k$, $k \geq 3$
encode the corrections to the superpotential involving
non-renormalizable higher dimensional couplings.

\TABLE{\renewcommand{\arraystretch}{1.7}
\begin{tabular}{ccc} 
\hline 
\hline
$A^\infty$ category & Fukaya Category & Intersecting D-branes  \\ 
\hline 
\hline
\vspace{-0.25cm}
Objects & special Lagrangian submanifolds $\Pi_\a$ & D-brane stacks \\
& endowed with unitary systems $\ce_\a$ & with flat gauge bundles \\
\hline
\vspace{-0.25cm}
Morphisms &  $Hom(\ce_\a, \ce_\b)$ generated by & 
Chiral matter in bifundamentals \\
& intersection points $p_{\a\b} \in \Pi_\a \cap \Pi_\b$ & 
localized at the intersections\\
\hline
\vspace{-0.25cm}
Grading & Maslov index $\eta(p_{\a\b})$ & Spectral flow index \\
& & (mod $2$ $=$ chirality) \\
\hline
Composition maps & Holomorphic discs & 
Worldsheet instanton corrections \\
& $m_1$ & mass operator \\
& $m_2$ & Yukawa coupling \\
\vspace{-0.25cm}
& $m_k$, $k \geq 3$ & non-renormalizable couplings \\
& & involving $k+1$ chiral fields \\
\hline
\end{tabular} 
\label{corresp}
\caption{\small Fukaya-Yukawa dictionary.}}

We have summarized in table \ref{corresp} the physical interpretation of the
basic quantities conforming Fukaya's category. This list is not
meant to be complete. In fact, it turns out that Fukaya's categories
are not always well defined, and not until very recently has 
a complete definition of such construction been provided \cite{Fu12}. 
The computation of such category in the simplest case of a Calabi-Yau
one-fold, i.e., the elliptic curve, was performed in \cite{poli}, where
the match with the derived category of coherent sheaves in a T-dual torus
was also performed, proving Kontsevich's conjecture in this case.
The extension to higher-dimensional symplectic tori followed in
\cite{theta}. The similarities of the results in those papers with the 
computations of section 3 are not an accident, but a sample
of the deep mathematical meaning of the computation of Yukawa couplings
in intersecting D-brane models.

\section{Final comments and conclusions}

In the present paper we have addressed the computation of
Yukawa couplings among chiral fields living at the intersections
of wrapping D-branes. The simplest class of models considered
are toroidal (orientifold) compactifications of Type IIA string
theory with D6-branes wrapping factorizable 3-cycles on 
$T^2 \ti T^2 \ti T^2$.
The Yukawa couplings come from holomorphic world-sheet  
instanton contributions. We find that the Yukawa couplings 
have simple expressions in terms of (products) of complex
Jacobi theta functions with characteristics. They depend on the
K\"ahler class of the tori and on the open string moduli but not
on the complex structure. They do not depend explicitly on the particular 
wrapping numbers of the D6-branes but rather on the intersection numbers
on the different subtori. 

This class of toroidal models are of phenomenological interest since 
one can find
specific D-brane configurations with the chiral spectrum of the SM 
living at the intersections. We provide an specific example in which
the massless chiral spectrum is that of the minimal supersymmetric 
standard model and compute the corresponding physical Yukawa couplings.
They have a simple expression in terms of a product of two complex
theta functions depending on the K\"ahler class and the open string moduli.
In this simple model only one generation of quarks and leptons get
mass and the other two are massless. This may be considered 
a  satisfactory starting point for  a fully successful description of the
observed spectrum of quarks and leptons since  small deviations of the simple
configuration considered  could give rise to smaller masses for the
rest of the generations. A more complete study of the structure of
 Yukawa couplings
in intersecting D-brane models is of clear interest and is at present underway
\cite{cim6}. 

The methods we employed for the computation of Yukawa couplings may also 
be applied to more general manifolds. In particular we have shown how
one can compute the Yukawa couplings in certain classes of 
elliptically fibered CY manifolds  which are mirror to D3-branes 
sitting on complex cones over  del Pezzo surfaces.
We compute the  Yukawa couplings on a particular simple manifold
which is known to be mirror to D3-branes sitting on  a ${\cpx^3}/\inte_3$ 
singularity and show that they match with the known Yukawa 
couplings in the mirror theory.
Similarly, one could foresee computing Yukawa couplings of more 
complicated models involving other del Pezzo surfaces, as
well as those models which involve D-branes with magnetic fluxes
on Type IIB or Type I toroidal compactifications \cite{bgkl,aads}.
The latter are related with our intersecting D-brane configurations of
section 3 by plain toroidal T-duality \cite{bdl,torons}.

We have  found a connection between Yukawa couplings 
in D6-brane intersecting brane setups of Type IIA string theory 
and Fukaya's  category in a symplectic manifold. 
We provide a dictionary giving the mathematical concept associated
to each of the physical objects. 
It would be
very interesting to pursue this connection further and provide
a more detailed dictionary.

In summary, we have shown how the computation of Yukawa couplings 
in intersecting D-brane configurations 
offers an interesting point of connection of very phenomenological 
questions like the structure of quark and lepton masses and 
very abstract mathematical notions like that of Fukaya's 
category.
We find amusing that, after all, the work of Yukawa could
be connected to that of Fukaya!

\bigskip

\bigskip

\bigskip

\centerline{\bf Acknowledgments}

\bigskip

We are grateful to G. Aldaz\'abal, F. Quevedo, R. Rabad\'an and  A. Uranga 
 for very
useful  comments and discussions.
L.E.I and F.M. thank CERN's Theory Division were part of 
this work was carried out.
The research of D.C. and F.M. was supported by
 the Ministerio de Educaci\'on, Cultura y Deporte (Spain) through FPU grants.
This work is partially supported by CICYT (Spain) and the
European Commission (RTN contract HPRN-CT-2000-00148).

\newpage

\appendix

\section{Higher dimensional holomorphic discs}

The aim of this section is to show the existence and uniqueness of 
(anti)holomorphic worldsheet instantons that connect factorizable
$n$-cycles in $T^{2n}$. Contrary to Euclidean intuition, these 
volume-minimizing surfaces are not, in general, given by flat triangles 
of $\real^{2n}$ but by more complicated calibrated manifolds.

Let us first describe the problem more precisely. For simplicity, instead
of dealing with the complex manifold $T^2 \ti \dots \ti T^2$ equipped 
with a complex structure, let us consider its covering space, given by 
$\cpx^n$. The factorizable $n$-cycles of (\ref{vectorscpx}) are mapped 
to affine Lagrangian $n$-planes given by 
\beq
L_\a  = \bigotimes_{r = 1}^n
\{ t^{(r)} \cdot z_\a^{(r)} + v_\a^{(r)} | 
t^{(r)} \in \real \},
\label{affine}
\eeq
where the fixed quantities $z_\a^{(r)}, v_\a^{(r)} \in \cpx$ define
the affine equation of a line in the $r^{th}$ complex plane. Of course,
we should also consider all the copies of such affine subspace under
$T^{2n}$ lattice translations, which implies the modification
$v_\a^{(r)} \mapsto v_\a^{(r)} + l_1^{(r)} + \tau^{(r)} l_2^{(r)}$, 
$l_1^{(r)}, l_2^{(r)} \in \inte$ in (\ref{affine}) 
(see figure \ref{tri} for an example of this). For our purposes, however, 
it will suffice to consider just a single copy given by (\ref{affine}).

Let us now consider a triplet of such $n$-hyperplanes $(L_a, L_b, L_c)$ in
$\cpx^n$. Two by two, these hyperplanes will either intersect at a single
point in $\cpx^n$ either be parallel in $s$ complex dimensions. 
If that is the case, we will consider that they are on top on each other 
on such complex dimensions, intersecting in a Lagrangian hyperplane
of real dimension $s < n$. We can visualize such geometry by projecting
it into each complex dimension $\cpx_{(r)}$. 
This picture will look as three real lines 
$l_a^{(r)}, l_b^{(r)}, l_c^{(r)} \subset \cpx_{(r)}$ 
forming a triangle whose vertices are the projection of the 
hyperplanes' intersections. If two hyperplanes are parallel
in such complex dimension, the corresponding triangle will be
degenerate (see figure \ref{multimap}).

A worldsheet instanton will be described by a map $\vphi : D \raw \cpx^n$
with the following properties:
\begin{itemize}

\item $\vphi$ is either holomorphic or antiholomorphic.

\item 
$\vphi(\om_{ab}) = L_a \cap L_b$,\quad
$\vphi(\om_{bc}) = L_b \cap L_c$,\quad
$\vphi(\om_{ca}) = L_c \cap L_a$,
where $\om_{ab}, \om_{bc}, \om_{ca} \in \p D$
are counterclockwise ordered in $\p D$.

\item 
$\vphi(\p D_{a}) \subset L_a$, \quad
$\vphi(\p D_{b}) \subset L_b$, \quad
$\vphi(\p D_{c}) \subset L_c$,
where $\phi(\p D_{a})$ is the part of $\p D$ between
$\om_{ca}$ and $\om_{ab}$, etc.

\end{itemize}
Furthermore, since we are only interested in the embedded surface $\vphi(D)$,
we must quotient our space of solutions by $Aut(D) = PSL(2,\real)$.

%
\EPSFIGURE{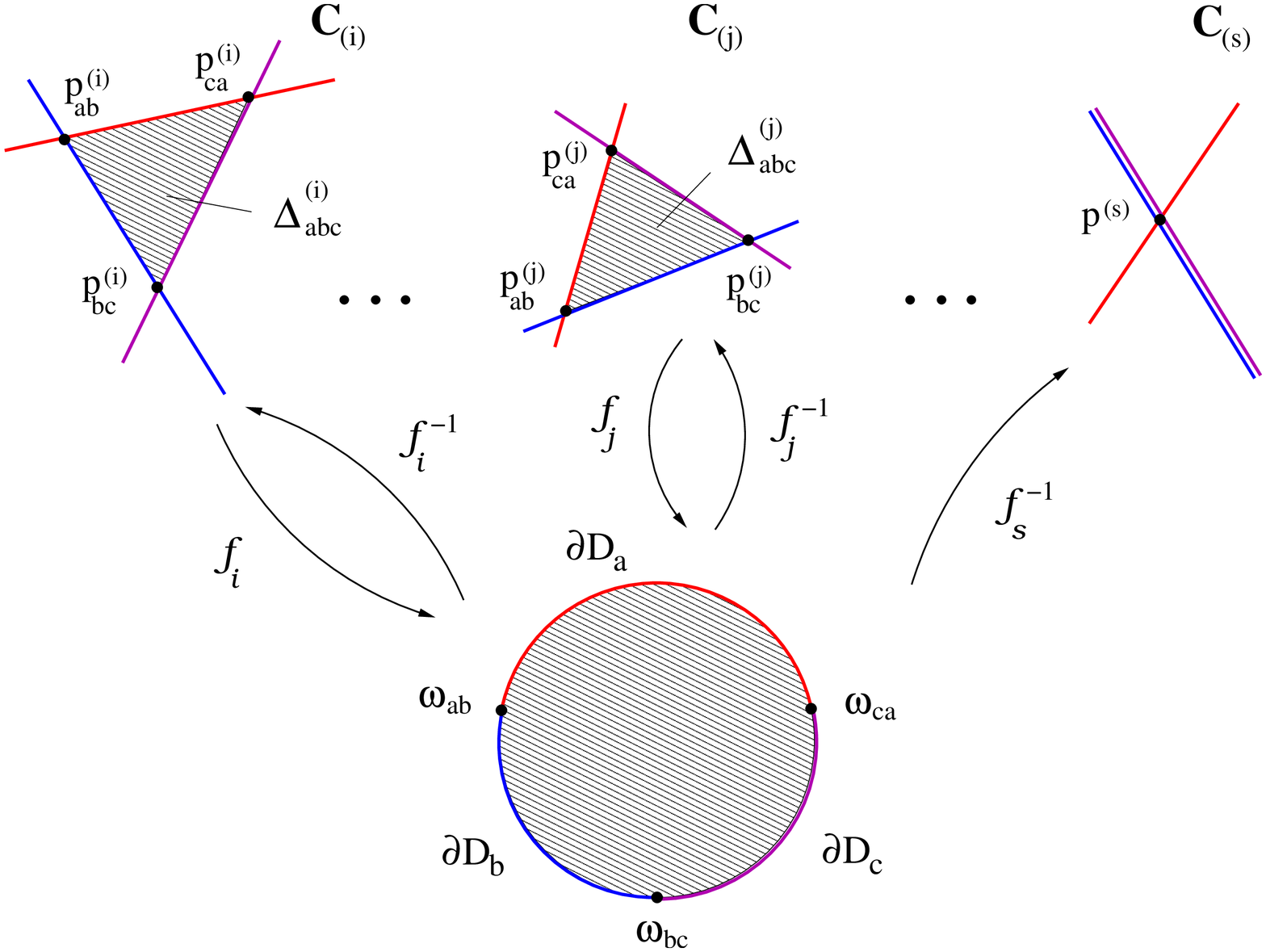, width=6in}
{\label{multimap}
Construction of higher dimensional holomorphic discs.}
%

Let us first focus on the simple case $n=1$, where the worldsheet instanton
will have the form of a triangle $\D_{abc} \subset \cpx$. 
The Riemmann Mapping Theorem \cite{conway} asserts that there exist
a one-to-one analytic function $f : \cpx \raw D$ such that the image of
$Int(\D_{abc})$ is $Int(D)$. Moreover, this function is unique up to
the group of conformal maps of the open unit disc into itself, which
is given by 
\beq
{\rm Conf}(D) = \{\phi_\om^\tau(z) = \tau \cdot {z - \om \over 1 - \bar  \om z}
| \om \in Int(D), \tau \in \p D \}.
\label{confdisc}
\eeq
A several statement holding for antianalytic functions. 
If we also require the boundary of such regions to match under $f$, that
is, if we require $f(\p \D_{abc}) = \p D$ following the second point above,
then we will select one unique map from the whole family 
$\phi_\om^\tau \circ f$, since Con$(D) \cong PSL(2,\inte)$,
and this latter group can be fixed by specifying the location of
$\om_{ab}$, $\om_{bc}$ and $\om_{ca}$ in $\p D$.
Moreover, the map $f$ will be either analytic or antianalytic,
but not both. Finally, the third point is satisfied by continuity of $f$.

We have thus shown that, for the case $n = 1$ there exist a unique
function, given by $f^{-1}$, that describes our worldsheet instanton. 
From this simple result we can derive the statement for general $n$. 
Indeed, let us consider the $n$ functions $f_{(r)}$ which are obtained 
from the above construction, now with the triangles $\D_{abc}^{(r)}$ 
which are defined from the lines 
$l_a^{(r)}, l_b^{(r)}, l_c^{(r)} \subset \cpx_{(r)}$. That is,
we consider the same problem for each projection into $\cpx_{(r)}$.
In case the triangle is degenerate (e.g., $l_b^{(s)} = l_c^{(s)}$
for some $s$) 
then such function does not exist, and we take $f^{-1}$ to be the 
constant map from $D$ to the common intersection point of the three
lines. After this, is easy to see that the desired (anti)holomorphic 
surface is given by $\phi(D) = \vec f^{-1}  (D)$, where
\beq
\vec f^{-1}(\om) = \left(f_1^{-1}(\om), f_2^{-1}(\om), 
\dots, f_n^{-1}(\om) \right) \in \cpx^n,  \quad \quad \om \in D,
\label{surface}
\eeq
and that this surface satisfies all the three requirements above.

A consequence of the holomorphic properties of $\vec f$ is that
 $S = \phi(D)$ will be a surface calibrated by the K\"ahler 2-form.
It is natural to wonder which is the 'shape' of $S$ in terms of 
$\cpx^n$ geometry. A na\"{\i}ve guess would lead us to think
that, being surface minimizing, it has a triangular shape.
This will not, however, be the generic case. Indeed, let us consider 
the relatively simple case of two-complex dimensions. In this case,
$\vec f^{-1} = \left(f_1^{-1}, f_2^{-1} \right)$ maps $D$ into
$\D_{abc}^{1} \times \D_{abc}^{2} \subset \cpx^2$. We can see
the surface $\vec f^{-1}  (D)$ embedded on $\cpx^2$ by considering 
the graph
\beq
(z_1, z_2) = (z_1 , f_2^{-1} \circ f_1 (z_1)), \quad z_1 \in  \D_{abc}^{1}.
\label{graph}
\eeq
Clearly, $f_2^{-1} \circ f_1 (\D_{abc}^{1}) = \D_{abc}^{2}$. 
This surface will be a triangle in $\cpx^2$ only if 
$f_2^{-1} \circ f_1$ is a constant complex number, that is,
only if $\D_{abc}^{1}$ and $\D_{abc}^{2}$ are congruent triangles.
A similar argument can be carried out for higher dimensions.

\end{document}